\documentclass[journal=jacsat,manuscript=article,layout=twocolumn]{achemso}

\usepackage{geometry}

\setkeys{acs}{articletitle=true}
\usepackage{lettrine}

\usepackage[version=3]{mhchem} % Formula subscripts using \ce{}
\usepackage[T1]{fontenc}       % Use modern font encodings

\usepackage[fontsize=10pt]{scrextend}

\usepackage{newtxtext}
\usepackage{bm} 
\usepackage[labelfont=bf]{caption} % bold ``Figure''

\usepackage[hyperfootnotes=false]{hyperref}	 % make refs, cites, links in color
\hypersetup{
  colorlinks,
  citecolor=blue,
  linkcolor=blue,
  urlcolor=blue
}

\usepackage{titlesec} % inline subsections
\titleformat{\subsection}[runin]
{\normalfont\bfseries}{\thesubsection{.}}{1em}{}[.]

\titleformat{\section}
{\normalfont\sffamily\large\bfseries}{\thesection{.}}{1em}{\MakeUppercase}

\usepackage[table]{xcolor}
\usepackage{graphicx}
\usepackage{amsmath,fullpage,amssymb}
\usepackage{graphics,epsfig}
\usepackage{makeidx}
\usepackage{gensymb}
\usepackage{siunitx}
\usepackage{subcaption}

\usepackage{soul}	% for strikethrough font

\def\ln{{\operatorname{ln}}}

\def\rme{{\mathrm{e}}}

\def\w{\mathrm{w}}
\def\cav{\mathrm{cav}}

\def\cap{\mathrm{cap}}

\def\Eq{eq.}
\def\Eqs{eqs.}

\def\Fig{\textcolor{blue}{Figure}}
\def\Figs{\textcolor{blue}{Figures}}
\def\Tab{\textcolor{blue}{Table}}

\newcommand{\kB}{k_\mathrm{B}}

\newcommand{\trm}[1]{{\textrm{#1}}}

\SectionNumbersOn
\let\oldmaketitle\maketitle
\let\maketitle\relax

\title{\flushleft Nanoscale defects and heterogeneous cavitation in water
}

\author{Marin \v{S}ako}
\affiliation{\rm\small Department of Theoretical Physics, Jo\v{z}ef Stefan Institute, 1000 Ljubljana, Slovenia}
\alsoaffiliation{\rm\small Faculty of Mathematics and Physics, University of Ljubljana, 1000 Ljubljana, Slovenia}

\author{Fabio Staniscia}
\affiliation{\rm\small Department of Theoretical Physics, Jo\v{z}ef Stefan Institute, 1000 Ljubljana, Slovenia}

\author{Roland R.\ Netz}
\affiliation{\rm\small Fachbereich Physik, Freie Universit\"{a}t Berlin, Berlin 14195, Germany }

\author{Emanuel Schneck}
\affiliation{\rm\small Physics Department, Technische Universität Darmstadt, 64289 Darmstadt, Germany }

\author{Matej Kandu\v{c}}
\affiliation{\rm\small Department of Theoretical Physics, Jo\v{z}ef Stefan Institute, 1000 Ljubljana, Slovenia}
\email{matej.kanduc@ijs.si}

\begin{document}
\pagenumbering{arabic}
\noindent

\parindent=0cm
\setlength\arraycolsep{2pt}

\twocolumn[	% make wide abstract
\begin{@twocolumnfalse}
\oldmaketitle

\begin{abstract}\small
Cavitation, the formation of vapor bubbles in metastable liquids, is highly sensitive to nanoscale surface defects. Using molecular dynamics simulations and classical nucleation theory, we show that pure water confined within defect-free walls can withstand extreme negative pressures, far beyond those observed experimentally. Hydrophobic surfaces trigger heterogeneous cavitation and lower the cavitation pressure magnitude, but not to experimental levels.
Notably, a single nanoscopic surface defect capable of hosting a vapor bubble drastically reduces the tensile strength of water. We find that defects as small as 1--2 nm can act as effective cavitation nuclei, a scale smaller than predicted by simple mechanical stability arguments. This discrepancy arises from stochastic fluctuations of the vapor bubble, which can overcome the kinetic free-energy barrier for cavitation. Our findings show that cavitation is predominantly determined by the largest surface defect rather than the overall defect density, emphasizing the importance of eliminating the largest surface imperfections to enhance stability against cavitation.

\vspace{5ex}
\end{abstract}
\end{@twocolumnfalse}]

\maketitle
\setlength\arraycolsep{2pt}
\small

\section{Introduction}
When a liquid is subjected to pressures below its saturated vapor pressure, it enters a metastable state where it becomes susceptible to the formation of vapor bubbles, a phenomenon known as cavitation. Under certain conditions, the liquid can endure tensile deformation for extended periods of time, effectively becoming ``stretched'' under negative pressure. The study of stretched water dates back to the 17th century~\cite{caupin2006cavitation}, and it continues to captivate researchers today because of its scientific and technological relevance~\cite{stroock2014physicochemical,wheeler2008transpiration, pagay2014microtensiometer, vincent2014drying, dular2004relationship, reuter2022cavitation, adhikari2015mechanism, peters2015numerical}, as well as the persistent discrepancies between theory and experiment~\cite{caupin2006cavitation, herbert2006cavitation, caupin2013stability}.

While classical nucleation theory (CNT) suggests that water should cavitate at pressures ranging between $-$120 and $-$160 MPa~\cite{fisher1948fracture, caupin2005liquid, caupin2006cavitation, caupin2013stability, azouzi2013coherent},  experiments involving macroscopic volumes have consistently fallen short of this theoretical threshold.
In unpurified water, cavitation typically occurs close to the saturated vapor pressure. Only when water is thoroughly degassed, purified, and contained within vessels with hydrophilic surfaces does the cavitation pressure decrease significantly and becomes negative. Even under these conditions, it only reaches about $-$30 MPa at best~\cite{briggs1950limiting, caupin2006cavitation, caupin2013stability, caupin2015escaping}.
%A notable exception involves microsized water inclusions in quarz minerals~\cite{zheng1991liquids, alvarenga1993elastic, azouzi2013coherent}, where negative pressures up to $-$140 MPa have been observed, though the underlying reasons for this discrepancy remain poorly understood.

It is now widely recognized that nucleation rarely occurs within the bulk of a homogeneous liquid. Instead, it predominantly initiates at various heterogeneities in the system, such as solid walls, gas bubbles, dissolved particles, and other impurities~\cite{herbert2006cavitation, morch2007reflections, gao2021effects}. 
These heterogeneities, known collectively as {\sl cavitation nuclei}, play a crucial role in the inception of cavitation through a process referred to as {\sl heterogeneous nucleation}. 
Cavitation nuclei are typically long-lived and, in most practical scenarios, contain at least some amount of gas~\cite{jones1999bubble, morch2007reflections}.
Because of their inherent instability against dissolution, gas bubbles are not expected to remain freely suspended within the liquid~\cite{epstein1950stability, lohse2015surface, tan2021stability}. However, they can become stabilized when they attach to solid surfaces or reside within crevices~\cite{harvey1944bubble, jones1999bubble, morch2007reflections, lohse2015surface, petsev2020universal}. These surfaces can include the bounding walls of the vessel or the surfaces of suspended solid particles within the liquid~\cite{atchley1988thresholds, madanshetty1991acoustic}.

The idea that pre-existing gas-filled crevices serve as cavitation nuclei was already proposed by Harvey et al.\ in 1944~\cite{harvey1944bubble} and refined theoretically by Atchley and Prosperetti in 1989 to account for the loss of {\it mechanical stability} of the nucleus~\cite{atchley1989crevice}.
While this theory has been qualitatively supported by numerous experiments, it was not until 2009 that a quantitative validation was provided by Borkent et al.~\cite{borkent2009nucleation}. Using fabricated pits with precisely controlled diameters down to 100 nm, they demonstrated that these artificial defects could reliably initiate cavitation, with results closely aligning with theoretical predictions based on mechanical stability. 

However, as surface defects and the bubbles they host shrink to the nanoscale, a fundamental question arises: Do the mechanisms of cavitation change at these scales? While nanometer-sized bubbles may be mechanically stable against expansion, they could be {\it kinetically} unstable. In other words, thermal fluctuations may overcome the free-energy barrier required for cavitation. To account for this, kinetic theories such as CNT are more appropriate than mechanical stability arguments.
An intriguing question also concerns the critical defect size capable of acting as a cavitation nucleus.
Nanoscale defects and bubbles are practically unavoidable. Advanced imaging techniques have shown that even highly polished surfaces contain nanoscale features like bumps, steps, and cavities~\cite{de2013surface, siretanu2016atomic}. Hydrophobic nanoscale cavities are particularly noteworthy, as they can be spontaneously dewetted~\cite{lum1999hydrophobicity, huang2003dewetting, sharma2012evaporation}, forming vapor nanobubbles even in the absence of additional gas. This phenomenon highlights the potential role of hydrophobic nanoscale defects in cavitation.
The experimental challenges of observing and quantifying nanoscale phenomena leave significant gaps in our understanding of how these defects influence cavitation. 

To address these questions, we employ Molecular Dynamics (MD) simulations to investigate cavitation initiated at nanoscale pits on flat surfaces in water. 
By focusing on these tiny imperfections, we aim to provide a detailed understanding of how nanoscale surface defects (specifically cylindrical pits) influence cavitation, offering insights beyond the concept of mechanical stability.

\section{Methods}
\subsection{Simulation model}
In our simulations, we use the TIP4P/2005 water model~\cite{tip4p_water}, a 4-point water model known for its accuracy in reproducing experimental interfacial tensions and for its compatibility with the CHARMM36 force field. 
To model atomistically smooth surfaces, we employ a self-assembled monolayer (SAM) described by the CHARMM36 force field~\cite{vanommeslaeghe2010charmm} that was used in our previous study~\cite{10.1093/pnasnexus/pgad190}. The SAM consists of ten-carbon alkyl chains terminated by hydroxyl (OH) groups (i.e., $n$-decanol molecules), arranged in a hexagonal lattice with a nearest-neighbor distance of 0.497 nm, mimicking a SAM on a gold surface, Au(111)~\cite{strong1988structures,chidsey1990chemical}. The molecules are tilted by 30° relative to the surface normal~\cite{fenter1997epitaxy}. 
The SAM structure is stabilized by harmonic restraining potentials applied at two positions: the second carbon atom from the OH group is restrained with a force constant of $k_x = k_y = k_z = 300$ kJ mol$^{-1}$ nm$^{-2}$, and the terminal carbon atom (in the CH$_3$ group) is restrained with a force constant of $k_x = k_y = k_z = 500$ kJ mol$^{-1}$ nm$^{-2}$.

We tune the surface dipole strength and contact angle by scaling the partial charges of the OH group and its three nearest carbon neighbors. For hydrophilic surfaces, we scale the charges by 0.8, and for hydrophobic surfaces, we set the charges to zero. 
The contact angles for each surface type are measured using the sessile droplet method, where cylindrical droplets of various sizes are simulated on the substrate, as was done in our recent study using a different water model~\cite{10.1093/pnasnexus/pgad190}. The macroscopic contact angles are then obtained by extrapolating the measured values to infinitely large droplets. Using this method, we determine the contact angles to be $\theta\approx 60\textrm{°}$ for our hydrophilic surfaces and  $\theta \approx 115\textrm{°}$ for the hydrophobic surfaces.

\subsection{Simulation details}
MD simulations are performed using Gromacs 2022.1~\cite{ABRAHAM201519}, with an integration timestep of 2 fs. 
Both electrostatic and Lennard-Jones (LJ) interactions are treated using the Particle--Mesh Ewald (PME) methods, with a real-space cutoff of 1.4 nm.
The system temperature is maintained at 300 K using the v-rescale thermostat~\cite{v-rescale}. 
Pressure is controlled using the C-rescale barostat~\cite{c-rescale}, with a compressibility value of $4.5 \times 10^{-5}$ $\mathrm{bar^{-1}}$.

\subsection{Pressure ramp method}
A cavitation pathway in a system at a fixed pressure, $p$, is quantified by its cavitation rate $k$, which represents the number of cavitation events per unit time. The inverse of this rate corresponds to the mean cavitation time, $\tau=k^{-1}$. 
According to the reaction rate theory, the cavitation rate is given by~\cite{hanggi1990reaction, Kanduc10733} 
\begin{equation}
\label{eq:k}
k = k_0 \exp{(-\beta G^*)}
\end{equation}
where $\beta=1/\kB T$, $\kB$ is the Boltzmann constant and $T$ the absolute temperature.  The kinetic prefactor $k_0$ denotes the frequency of cavitation attempts, and the Boltzmann factor accounts for the probability that an individual cavitation attempt overcomes the free energy barrier $G^*$ at pressure $p$.
 
While $G^*$ can be obtained from continuum approaches, simulations are required to determine $k_0$.  
However, assessing $k_0$ in simulations under constant negative pressure by measuring the mean cavitation time $\tau$, as described by \Eq~\ref{eq:k}, proves impractical. The primary challenge lies in the strong dependence of cavitation time on the applied pressure. 
Cavitation events only occur within a narrow range of negative pressures that permit feasible simulation times without reducing the free energy barrier too much. 

To overcome this obstacle, we employ the pressure ramp method~\cite{boucher2007pore, Kanduc10733, sako2024impact}, in which a linearly decreasing negative pressure over time (i.e., a linearly increasing magnitude) is applied, given by $p(t) = \dot p t$, where $\dot{p} < 0$ is the rate of pressure change. In this approach, $G^*$ gradually decreases over time, eventually reaching a low enough value to trigger cavitation within the simulation time. This approach makes the cavitation rate time-dependent. Solving the time-dependent transition rate equations for $k(t)$, as outlined in Ref.~\citenum{Kanduc10733}, leads to the following expression for the cavitation pressure,
\begin{equation}
    \label{eq:p_star}
    p^*_{\mathrm{cav}} = \dot p \int_0^\infty e^{-k_0 I(t)} \mathrm{d}t
\end{equation}
where
\begin{equation}
    \label{eq:integral}
    I(t) = \int_0^t e^{-\beta G^*(t')} \mathrm{d} t'
\end{equation}
with $G^*(t)$ being the time-dependent free energy barrier as a direct consequence of time-dependent pressure, $p(t)$. Equation \ref{eq:p_star} predicts the mean cavitation pressure, $p^*_{\mathrm{cav}}$, for a system subjected to a pressure ramp at the rate $\dot{p}$. It is important to distinguish this {\it dynamic} cavitation pressure $p^*_{\mathrm{cav}}$ (denoted with an asterisk) from the cavitation pressure $p_{\mathrm{cav}}$, which corresponds to the static pressure required for cavitation to occur within a time $\tau$. 

Since Gromacs does not natively support pressure-ramp simulations, we have devised an alternative approach by running a series of short simulations at a constant pressure~\cite{Kanduc10733}. In each successive simulation, the pressure is reduced by 0.1 MPa relative to the previous one, effectively simulating a linear pressure decrease. The duration of each simulation is determined by the desired pressure rate, $\dot p$. 
A cavitation event is identified by a sudden increase in the system’s volume, following a slow linear increase due to the decreasing pressure. To pinpoint the onset of the abrupt volume increase, we utilize the random sample consensus (RANSAC)~\cite{ransac, sako2024impact} algorithm. 

For each setup, we performed 6 to 8 independent runs at a given $\dot p$ to estimate the mean dynamic cavitation pressure, $p^*_{\mathrm{cav}}$. 
Finally, to obtain $k_0$, we fit \Eq~\ref{eq:p_star} to the simulated $1 / p_{\mathrm{cav}}^{*2}$ versus $-\dot p$ data, with $k_0$ as a fitting parameter.

%\subsection{Homogeneous cavitation in pure water}

\section{Free energy barriers for cavitation}

\begin{figure*}[h]\begin{center}
\begin{minipage}[b]{0.189\textwidth}\begin{center}
\includegraphics[width=\textwidth]{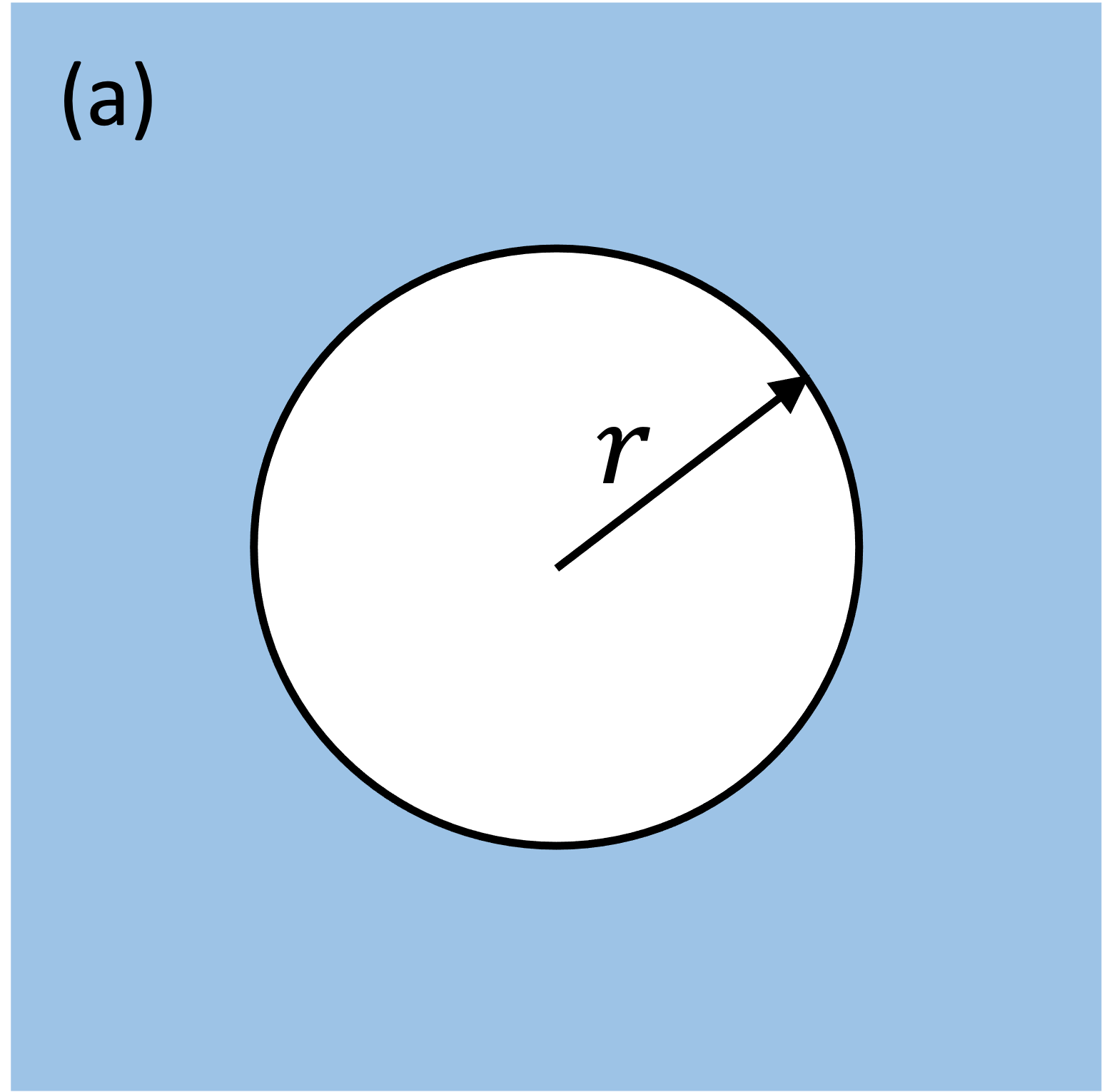}
\end{center}\end{minipage}
\begin{minipage}[b]{0.19\textwidth}\begin{center}
\includegraphics[width=\textwidth]{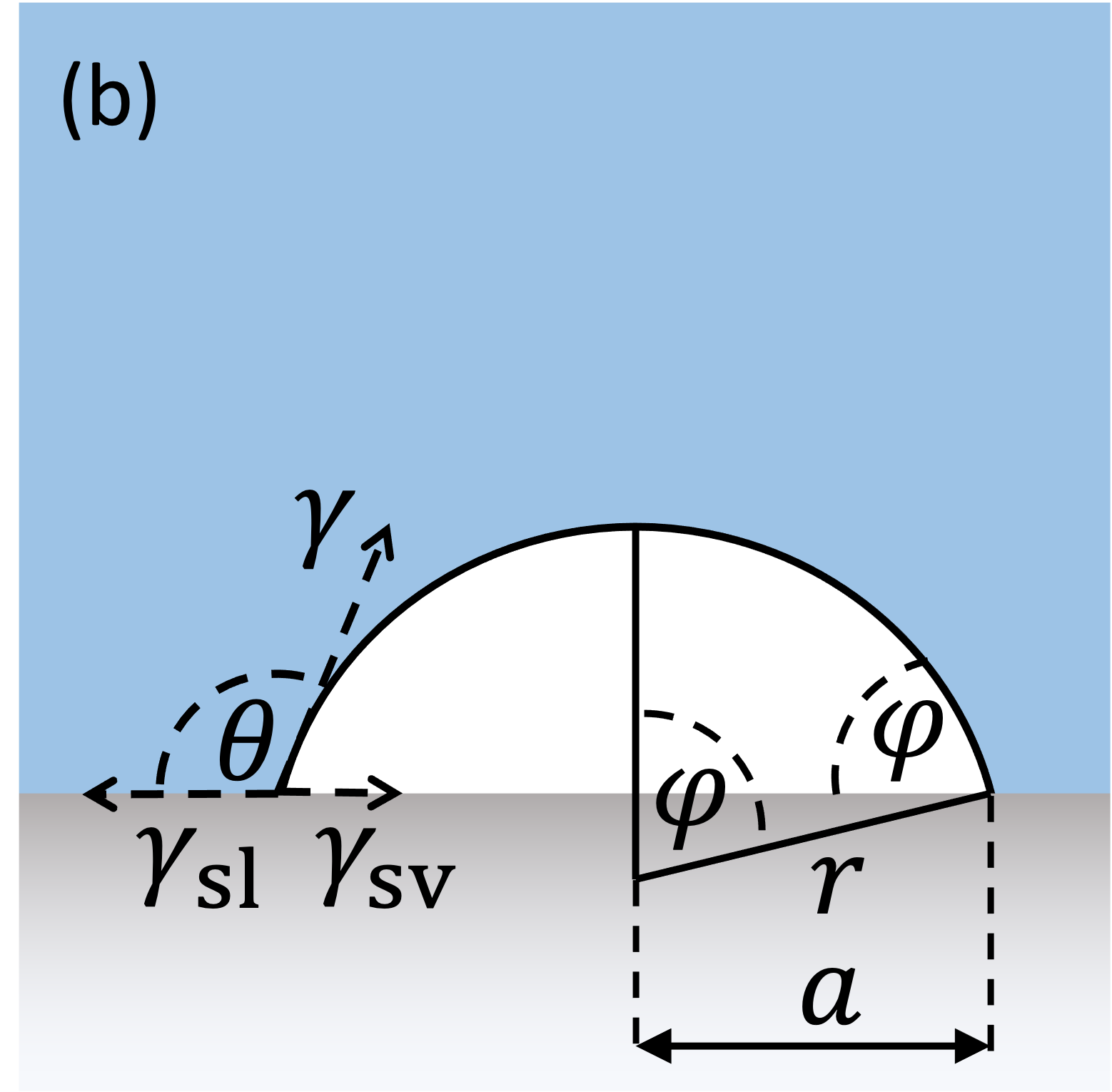}
\end{center}\end{minipage}
\begin{minipage}[b]{0.19\textwidth}\begin{center}
\includegraphics[width=\textwidth]{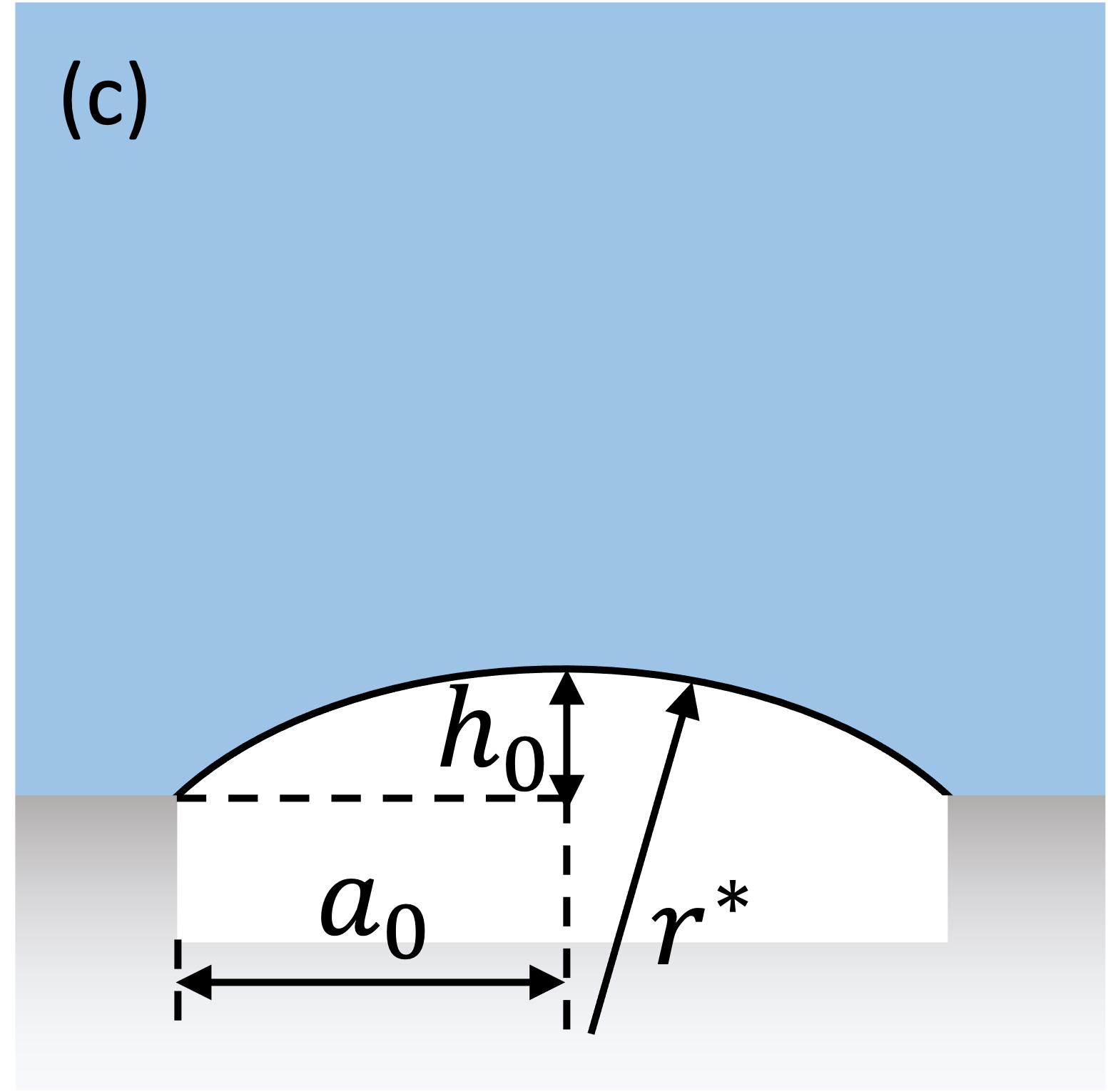}
\end{center}\end{minipage}
\begin{minipage}[b]{0.19\textwidth}\begin{center}
\includegraphics[width=\textwidth]{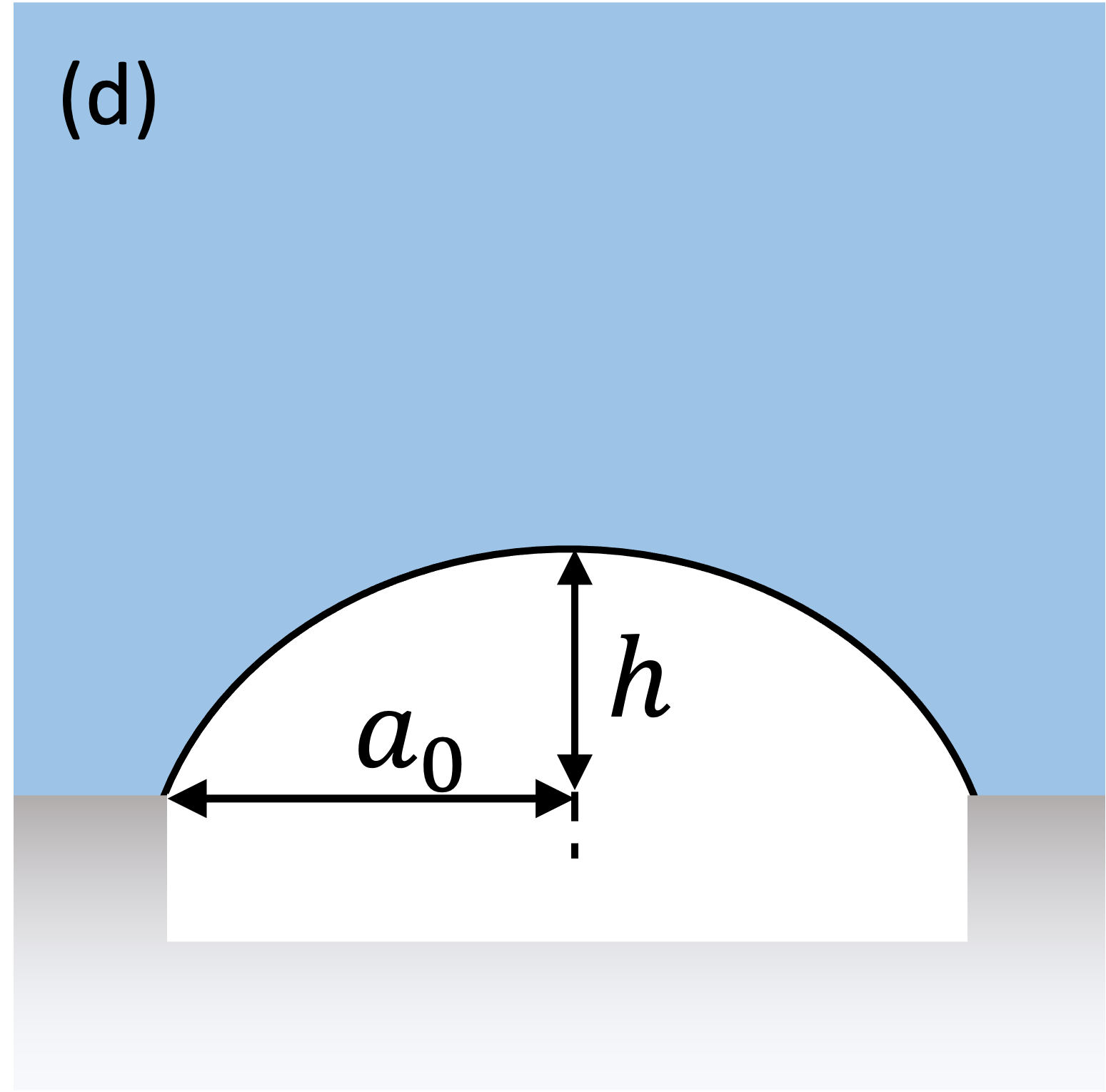}
\end{center}\end{minipage}
\begin{minipage}[b]{0.194\textwidth}\begin{center}
\includegraphics[width=\textwidth]{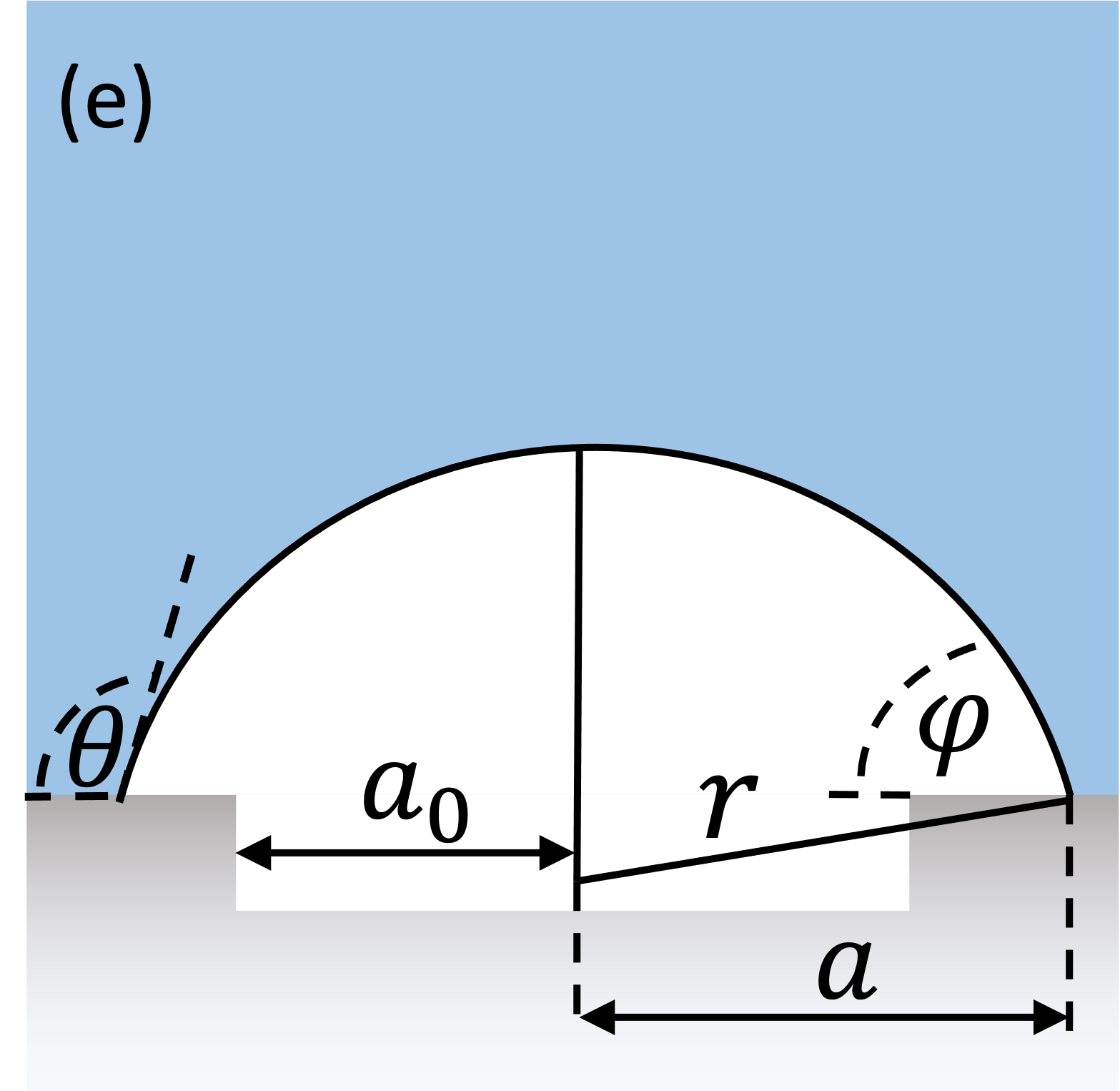}
\end{center}\end{minipage}
\caption{
Schematic illustration of nucleating bubbles:
(a) Homogeneous bubble nucleation in bulk liquid. 
(b) Heterogeneous bubble nucleation on an atomically smooth surface with a wetting contact angle $\theta$, where the bubble forms the supplementary contact angle $\varphi=180\textrm{°}-\theta$ with the surface.
(c) Metastable vapor bubble (i.e., in a local free energy minimum) in a hydrophobic pit with radius $a_0$, where the radius of its cap coincides with the critical radius  $r^*$.
(d) Expanding bubble with (i.e., out of equilibrium) a pinned contact line, valid as long as $\varphi<180\textrm{°}-\theta$. 
(e) Once the bubble unpins, it spreads over the surface with a contact angle $\varphi=180\textrm{°}-\theta$.
}
\label{fig:geometries}
\end{center}\end{figure*}

Before delving into simulating cavitation processes, we examine the free energies of the various nucleation pathways relevant to our study. We illustrate these pathways in \Fig~\ref{fig:geometries}, which include: (a) {\it homogeneous nucleation} in bulk water, (b) {\it heterogeneous nucleation} at a defect-free surface, and cavitation triggered by (c) a {\it preexisting vapor bubble} hosted by a pit in the surface. Depending on the contact angle, the bubble can expand in two distinct manners; (d) with a {\it pinned} contact line or (e) with its contact line {\it spreading} over the surface. The free energies associated with these pathways are essential for applying CNT. In CNT, cavitation inception occurs when a bubble stochastically forms and grows beyond a critical size, successfully overcoming the associated free-energy barrier~\cite{caupin2006cavitation, azouzi2013coherent, menzl2016molecular}.

\subsection{Homogeneous cavitation}
The primary and conceptually simplest form of cavitation is homogeneous cavitation, where a bubble forms within the bulk liquid.
The free energy associated with the expansion of a bubble of radius $r$, depicted in \Fig~\ref{fig:geometries}a, is given by
\begin{equation}
\label{eq:G_w}
G_\mathrm{w} = 4 \pi \gamma r^2 + \frac 43 \pi p r^3
\end{equation}
The first term represents the surface free energy required to create the bubble interface, where $\gamma$ is the liquid--vapor surface tension.
We will assume that $\gamma$ is independent of curvature to simplify the analysis. However, in our recent study~\cite{sako2024impact} we demonstrated that applying the Tolman-length correction to the surface tension introduces an adjustment to the cavitation pressure of approximately 20\%.
The second term represents the work performed by the negative pressure $p$ during volume expansion. We neglect the effect of the vapor inside the bubble, as typical cavitation pressures (tens of MPa) are much higher in magnitude than the saturated vapor pressure at room temperature (several kPa).

The competition between the two terms in \Eq~\ref{eq:G_w} results in a free energy maximum, $G^*_\w=G_\mathrm{w}(r^*)$, at the critical radius $r^*$. From the extremum condition, $\mathrm{d}G_\w / \mathrm{d} r = 0$, the critical radius is obtained as 
\begin{equation}
    r^* = -2 \gamma / p
    \label{eq:r_star}
\end{equation}
and the free energy barrier is 
\begin{equation}
    \label{eq:dG_star_w}
     G^*_\w = \frac{16 \pi \gamma^3}{3p^2}
\end{equation}
Notably, because of the cubic dependence of the energy barrier on surface tension, water exhibits an exceptionally high cavitation barrier for homogeneous nucleation~\cite{fisher1948fracture, caupin2005liquid, caupin2006cavitation, herbert2006cavitation, caupin2013stability, azouzi2013coherent, sako2024impact}.
As we will discuss in the following, homogeneous cavitation in water is virtually unattainable as it is superseded by other cavitation pathways with significantly lower barriers.

\subsection{Heterogeneous cavitation at a smooth surface}
The simplest form of heterogeneous cavitation occurs when a vapor bubble nucleates at a smooth, featureless surface~\cite{jones1999bubble}, as illustrated in \Fig~\ref{fig:geometries}b. 
In this case, the bubble expands over the surface at a fixed contact angle $\varphi$. 
Assuming quasi-equilibrium, the contact angle is supplementary to the wetting contact angle $\theta$ of the liquid, such that $\varphi + \theta = 180\textrm{°}$. The bubble free energy can be expressed as
\begin{equation}
    \label{eq:free_eng_het}
    G_\mathrm{s} = \gamma A_\cap  + (\gamma_\mathrm{sv}-\gamma_\mathrm{sl})A_\mathrm{base} + pV_\cap
\end{equation}
Here, $\gamma_\mathrm{sv}$ and $\gamma_\mathrm{sl}$ represent the solid--vapor and solid--liquid surface tensions, respectively. Using Young's equation, we  substitute $\gamma_\mathrm{sv}-\gamma_\mathrm{sl}=\gamma\cos\theta$.
The relevant geometric terms are as follows: 
$A_\cap = 2 \pi r^2 (1 - \cos \varphi)$ is the surface area of the spherical bubble cap, $A_\mathrm{base} = \pi a^2$ is its base area, and $V_\cap = (\pi/3) \, r^3 (2 - 3 \cos \varphi + \cos^3 \varphi)$ is its volume.
The critical bubble radius is determined by the condition $\trm{d} G_\mathrm{s} / \trm{d} r = 0$, giving again the critical radius in \Eq~\ref{eq:r_star}. The free energy barrier for heterogeneous cavitation at a smooth surface follows as
\begin{equation}
    \label{eq:free_eng_barr_het}
    G^*_\mathrm{s} = G^*_\mathrm{w} h(\theta)
\end{equation}
In this case, the free energy barrier equals the one for homogeneous cavitation, $G^*_\mathrm{w}$ (\Eq~\ref{eq:dG_star_w}), reduced by the geometric factor $h(\theta)$, given by~\cite{volmer1929keimbildung,caupin2006cavitation}
\begin{equation}
\label{eq:h}
h(\theta) = \frac 14(2 - \cos \theta) (1+\cos\theta)^2    
\end{equation}
Note that $h(\theta)$ is a monotonically decreasing function of $\theta$, starting from $h(0\textrm{°})=1$ for a completely wetting surface and approaching zero for a completely non-wetting surface $h(180\textrm{°})=0$. However, the
highest contact angles achievable for solid, atomistically smooth surfaces in water are around $\theta\approx 120$°~\cite{carlson2021hydrophobicity}, resulting in a barrier reduction of $h(120\textrm{°})\approx 0.16$. 
Despite this six-fold reduction compared to the barrier for homogeneous nucleation, it still remains considerably high, as we will see later on.

\subsection{Cavitation at a pit: Pinned bubble}

Consider a solid planar surface featuring a circular pit of radius $a_0$, as illustrated in \Fig~\ref{fig:geometries}c.
If the pit is deep enough and its walls sufficiently hydrophobic, water does not enter the pit~\cite{lum1999hydrophobicity}, resulting in the formation of a vapor bubble. Under negative pressure ($p<0$) that is not low enough to reach the mechanical stability limit (discussed later), this bubble is metastable and its liquid--vapor interface adopts the shape of a spherical cap with a radius  $r^*$. This radius is governed by the Laplace pressure and coincides with the critical radius (\Eq~\ref{eq:r_star}).

We denote quantities related to this metastable bubble by the superscript (0). The contact angle $\varphi^{(0)}$ of the bubble is given by $\sin\varphi^{(0)}=a_0/r^*$ (note that $\varphi^{(0)}<90$°).
For hydrophilic surfaces ($\theta<90$°), the condition for metastability is $r^*>a_0$, or equivalently $p/p_0<1$, where $p_0$ is defined as
\begin{equation}
    \label{eq:p_0}
    p_0 = - \frac{2 \gamma}{a_0}
\end{equation}
Thus, $p_0$ represents the limiting pressure for the mechanical stability of a bubble on hydrophilic surfaces ($\theta<90$°). 

For hydrophobic surfaces ($\theta>90$°), the condition is stricter, requiring that the contact angle remains smaller than the supplementary angle of the wetting contact angle, $\varphi^{(0)}<180\textrm{°}-\theta$, to prevent spreading over the surface. This translates to $\sin\varphi^{(0)}>\sin\theta$, or equivalently, $a_0/r^*>\sin\theta$. Substituting $r^*$ and using the definition of $p_0$, the metastability condition for hydrophobic surfaces becomes $p/p_0< \sin\theta$. 

When such a preexisting metastable bubble begins to expand, its base radius remains initially fixed, meaning the contact line is pinned, as depicted in \Fig~\ref{fig:geometries}d.
The corresponding free energy with respect to its metastable state can be expressed as
\begin{equation}
    \label{eq:free_eng_ws_2}
    G_\mathrm{pin} = \gamma(A_\cap  - A_\cap^{(0)}) + p (V_\cap - V_\cap^{(0)})
\end{equation}
Here, the first term represents the free energy associated with the increase in the bubble's cap surface area, while the second term accounts for the change in its volume. 

\begin{figure}\begin{center}
\begin{minipage}[b]{0.35\textwidth}\begin{center}
\includegraphics[width=\textwidth]{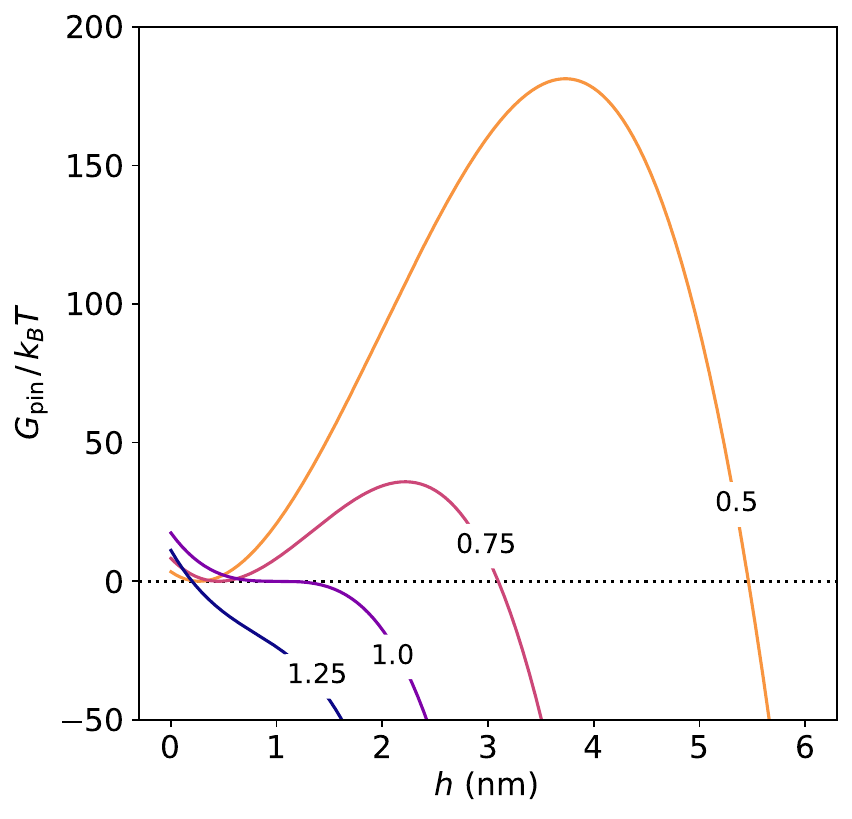}
\end{center}\end{minipage}
\caption{
Free energy of a pinned bubble in a pit with radius $a_0 = 1$ nm (as illustrated in \Fig~\ref{fig:geometries}d) as a function of the bubble height $h$, calculated using \Eq~\ref{eq:free_eng_ws_2}. 
The curves correspond to different values of $p / p_0$, as indicated by the labels.
For $p / p_0\ge 1$, the local free energy minimum vanishes, and the bubble is mechanically unstable.
}
\label{fig:polar_sam_defect_free_eng}
\end{center}\end{figure}

Since the bubble's radius of curvature changes non-monotonically during  expansion (i.e., decreases for $\varphi<90$° and increases for $\varphi>90$°), it is practical to use the height of the bubble, $h$, as the reaction coordinate.
In terms of $a_0$ and $h$, the surface area of the cap is given by $A_\cap = \pi (a_0^2 + h^2)$ and the volume by $V_\cap = \pi h (3 a_0^2 + h^2) / 6$.
This free energy of an expanding pinned bubble is shown in \Fig~\ref{fig:polar_sam_defect_free_eng} for various negative pressures.

For pressures above $p_0$ (i.e., $p/p_0<1$), the free energy curve exhibits a local minimum and a maximum, both satisfying $\trm{d} G_\mathrm{pin} / \trm{d} h = 0$.
The local minimum, occurring at a height $h^{(0)}=r^*-\sqrt{r^{*2}-a_0^2}$, corresponds to the initial metastable bubble (\Fig~\ref{fig:geometries}c), with a surface area of $A_\cap^{(0)} = A_\cap (h^{(0)})$ and a volume
 $V_\cap^{(0)} = V_\cap (h^{(0)})$ in \Eq~\ref{eq:free_eng_ws_2}.

As $h$ increases further, the free energy reaches the maximum at $h=h^*$, where the radius is again equal to $r^*$, just as in the local minimum.
At this critical bubble height $h^*=r^*+\sqrt{r^{*2}-a_0^2}$, free energy maximum,  $G^*_\mathrm{pin} = G_\mathrm{pin} (h^*)$, is given by
\begin{equation}
    \label{eq:G_star}
    G_\mathrm{pin}^* = G^*_\mathrm{w} K^3(p)
\end{equation}
where we introduced 
\begin{equation}
    \label{eq:K_of_p}
    K(p) = \sqrt{1 - \left( {p}/{p_0} \right)^2 }
\end{equation}
For small pit radii or low pressure magnitudes ($p/p_0\ll1$), $K(p)$ tends to 1, and the free energy barrier approaches that of homogeneous cavitation, provided that the bubble remains pinned. Conversely, when the pressure approaches the limiting value $p_0$, at which point the critical radius reaches the pit radius, $r^*=a_0$, the free energy barrier vanishes, as shown in \Fig~\ref{fig:polar_sam_defect_free_eng}. At this pressure, the system reaches the mechanical stability limit, beyond which the bubble becomes absolutely unstable and cannot be stabilized by the pit.

% when it unpins

The above free energy calculation applies to a bubble that remains pinned up to its critical size.
However, once its contact angle increases to $\varphi = 180\textrm{°} - \theta$, unpinning becomes energetically favorable, and the bubble subsequently spreads over the surface (as illustrated in \Fig~\ref{fig:geometries}e).
% This is true if no contact angle hysteresis occurs.

A key question, therefore, is whether unpinning occurs before the bubble reaches its critical size.
For pinned growth, the radius of curvature decreases until the contact angle reaches $\varphi = 90$°, after which the radius increases again. Thus, for hydrophobic surfaces ($\theta>90$°), the bubble unpins before $\varphi = 90$°, preventing it from reaching the critical size $r^*$ while pinned.  Only on hydrophilic surfaces ($\theta < 90^\circ$), the pinned bubble can reach the critical radius during the re-growth phase (for $\varphi > 90^\circ$).  
The pinning condition of a critical bubble on a hydrophilic surface $\varphi^* < 180\textrm{°} - \theta$ translates to $\sin\varphi^*>\sin\theta$, which leads to $a_0/r^*>\sin\theta$. Using \Eqs~\ref{eq:r_star} and \Eq~\ref{eq:p_0}, the pinning condition at criticality on hydrophilic surfaces expresses in terms of pressure as $p/p_0> \sin\theta$.

\subsection{Cavitation at a pit: Spreading bubble}
Once an expanding bubble in the pit unpins, it continues to spread over the surface with a growing base while maintaining a constant contact angle of $\varphi=180\textrm{°}-\theta$, as illustrated in \Fig~\ref{fig:geometries}e.
The free energy of the expanding bubble follows a similar expression as that for a bubble spreading on a flat, uniform surface (Eq.~\ref{eq:free_eng_het}), with the key difference being an offset due to the free energies associated with the initial metastable state, denoted by the superscript (0),
\begin{eqnarray}
    \label{eq:free_eng_ws_1}
    G_\mathrm{spr} &=& 
    \gamma(A_\cap- A_\cap^{(0)})  + \gamma (A_\mathrm{base} - A_\mathrm{base}^{(0)}) \cos \theta  
    \nonumber\\
    && + p( V_\cap- V_\cap^{(0)})
\end{eqnarray}
where the initial base area is the lateral area of the pit,  $A_\mathrm{base}^{(0)} = \pi a_0^2$, while $A_\cap^{(0)}$ and $V_\cap^{(0)}$ are again determined by the initial cap height $h^{(0)}$. 
The maximum of the free energy, satisfying $\trm{d} G_\mathrm{spr} / \trm{d} r = 0$, occurs at the critical radius $r^*$ and is given by
\begin{equation}
    \label{eq:free_eng_barr_spr}
    G^*_\mathrm{spr} = G^*_\mathrm{w} H(\theta,p)
\end{equation}
where we defined
\begin{equation}
    H(\theta,p) = \frac 14\left[2K(p) - \cos \theta\right] [K(p)+\cos\theta]^2   
\end{equation}
Notably, the function $H(\theta,p)$ is a generalization of the function $h(\theta)$, introduced in \Eq~\ref{eq:h}, for a pre-existing vapor bubble in a pit. In the limit of a vanishing pit ($a_0\to 0$), $K(p)$ approaches 1, and thus $H(\theta,p)\to h(\theta)$ is recovered.
\begin{figure}\begin{center}
\begin{minipage}[b]{0.35\textwidth}\begin{center}
\includegraphics[width=\textwidth]{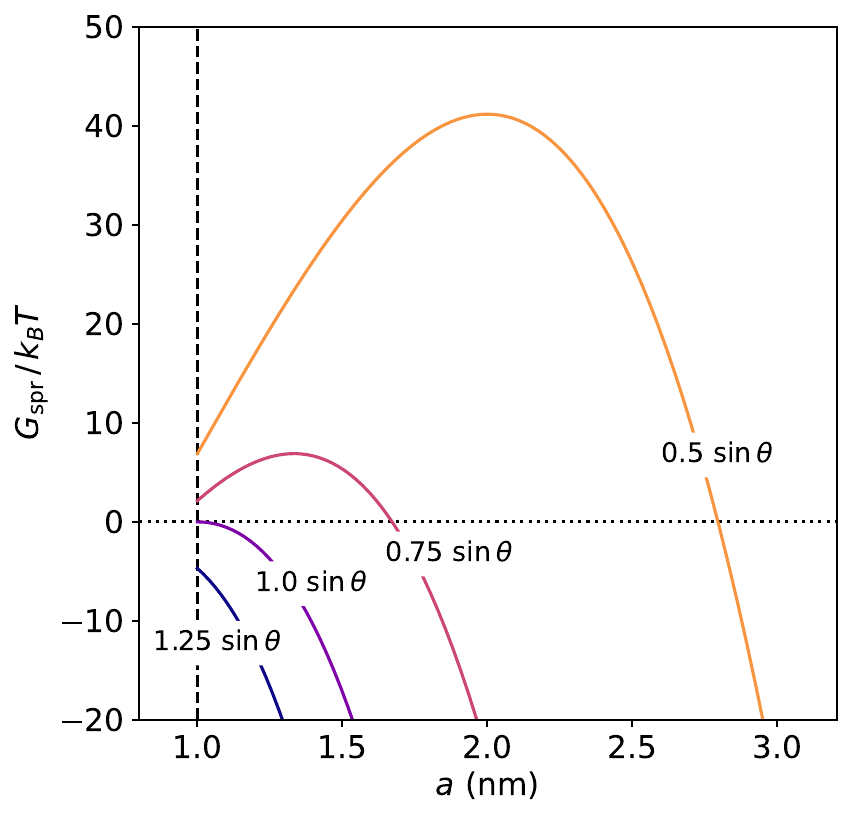}
\end{center}\end{minipage}
\caption{
Free energy of a bubble spreading over the surface from a pit with radius $a_0 = 1$ nm (as illustrated in \Fig~\ref{fig:geometries}e) as a function of the bubble base radius $a$, calculated using \Eq~\ref{eq:free_eng_ws_1} for a wetting contact angle of $\theta = 115\textrm{°}$. 
The curves correspond to different values of $p / p_0$, as indicated by the labels.
For $p / p_0\ge \sin\theta$, the local free energy minimum vanishes, and the bubble is mechanically unstable.
}
\label{fig:nonpolar_sam_defect_free_eng}
\end{center}\end{figure}

%For a hydrophobic surface, it unpins before reaching $\varphi=90\textrm{°}$, upon which the radius starts growing. Thus, the bubble on hydrophobic surfaces will always reach the critical bubble size in the unpinned case.

Finally, we can combine both modes of expansion---pinned and spreading---into a unified expression for the free energy barrier initiated from a preexisting metastable vapor bubble in a pit,
\begin{equation}
    G_\mathrm{pit}^*=
    \left\{\begin{array}{ll}
    G_\mathrm{spr}^* & \quad\textrm{for $p/p_0<\sin\theta$},\\
    G_\mathrm{pin}^* & \quad\textrm{for $p/p_0>\sin\theta$ and $\theta<90\textrm{°}$},\\
    0&\quad\textrm{for $p/p_0>\sin\theta$ and $\theta>90\textrm{°}$}.
  \end{array} \right.    
  \label{eq:G_star_pit}
\end{equation}
It is important to note that, as long as the bubble remains pinned when it reaches the critical size, the angle $\theta$ does not affect the free energy barrier (as given by \Eq~\ref{eq:G_star}).

\section{Results and discussion}
In a typical scenario involving negative pressure, a liquid is confined within a container with solid walls.
To model such a setup, we consider a cubic container with side length $L$, filled with water. 
We assume that the inner walls are atomistically smooth and characterized by a wetting contact angle $\theta$. Additionally, we assume that the edges and corners of the container are rounded, with radii of curvature substantially larger than the critical bubble radius $r^*$, to exclude additional nucleation sites.
Our objective is to investigate how preexisting vapor nanobubbles, hosted by hydrophobic defects on the container walls, influence the system stability under negative pressures.

\subsection{Container with defect-free surfaces}
We begin with the simplest scenario: a water-filled cubic container with defect-free surfaces and no pre-existing vapor bubbles. Under negative pressure conditions, cavitation can occur either within the bulk of the water (homogeneous cavitation) or at the water--surface interface (heterogeneous cavitation). The total cavitation rate is the sum of the rates for the two pathways~\cite{loche2024water}
\begin{align}\label{eq:k_comp}
    k = k_\mathrm{w} + k_\mathrm{s}
\end{align}
where $k_\mathrm{w}$ represents the cavitation rate within the bulk water, and $k_\mathrm{s}$ represents the cavitation rate at the surface.

The cavitation rate (see \Eq~\ref{eq:k}) within the bulk water volume
is given by~\cite{Kanduc10733}
\begin{equation}
    \label{eq:kw}
    k_\w = \kappa_\w V\,\rme^{-\beta G^*_\w}
\end{equation}
Here, the kinetic prefactor $\kappa_\w V$, representing the frequency of cavitation attempts, is proportional to the water volume, $V=L^3$, and $\kappa_\w$ denotes the attempt frequency density.
Similarly, the cavitation rate for heterogeneous cavitation at the solid walls is given by
\begin{equation}
    \label{eq:ks}
    k_\mathrm{s} = \kappa_\mathrm{s} A\,\rme^{-\beta G^*_\mathrm{s}}
\end{equation}
In the prefactor, $A=6L^2$ is the total surface area of the walls, and $\kappa_\mathrm{s}$ is the areal attempt frequency density, which generally depends on surface properties. In the exponent, $G^*_\mathrm{s}$ is the free energy barrier for heterogeneous cavitation, given by \Eq~\ref{eq:free_eng_barr_het}.

\begin{figure*}[h!]\begin{center}
\begin{minipage}[b]{0.135\textwidth}\begin{center}
\includegraphics[width=\textwidth]{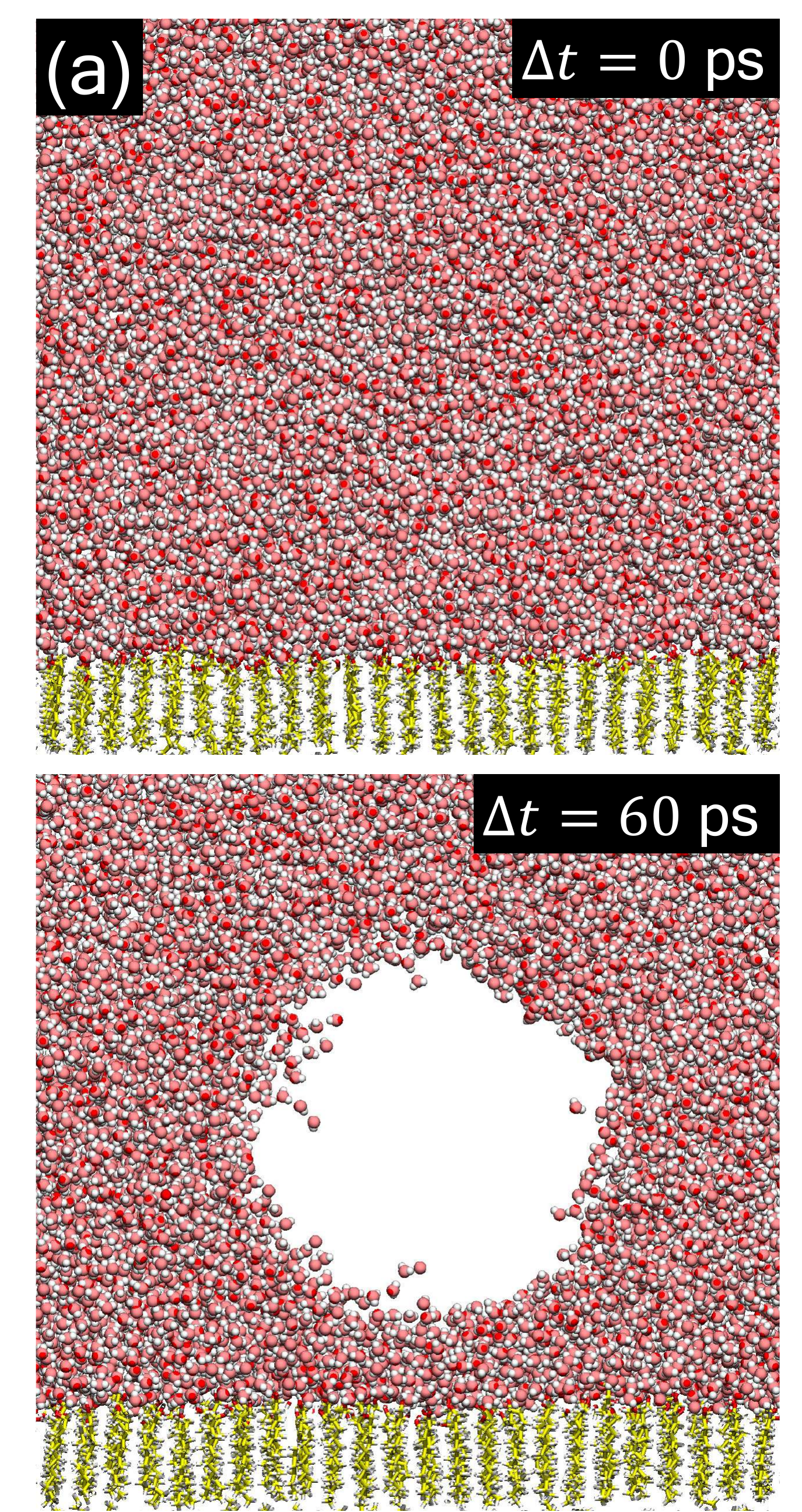}
\end{center}\end{minipage}
\begin{minipage}[b]{0.262\textwidth}\begin{center}
\includegraphics[width=\textwidth]{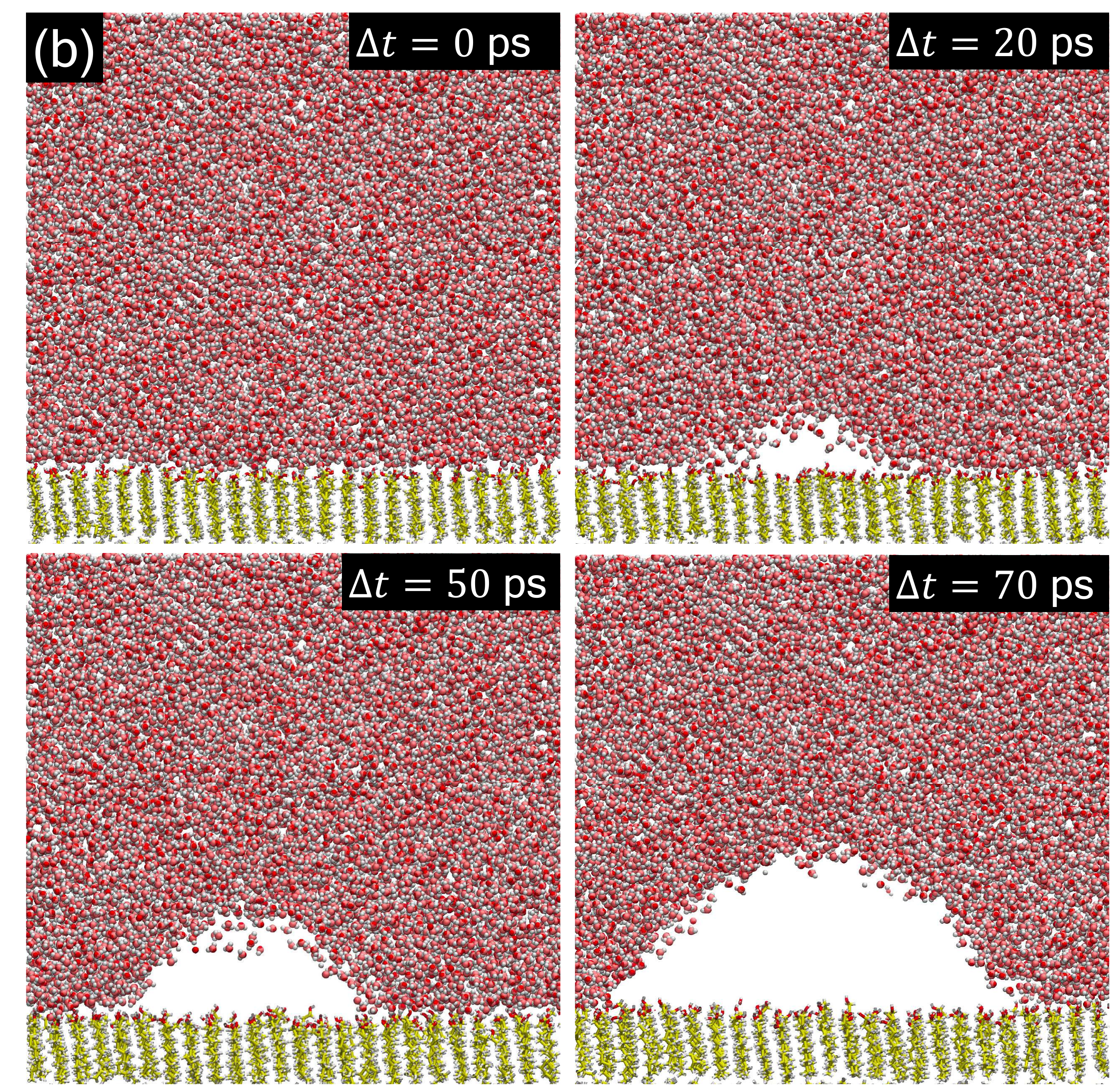}
\end{center}\end{minipage}
\begin{minipage}[b]{0.27\textwidth}\begin{center}
\includegraphics[width=\textwidth]{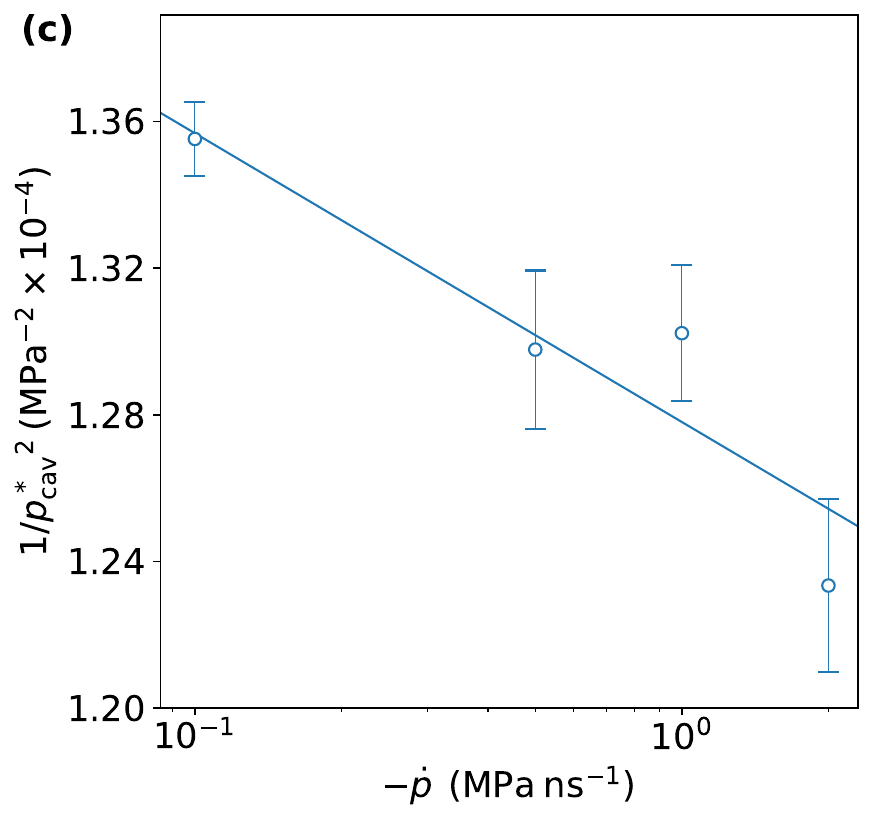}
\end{center}\end{minipage}
\begin{minipage}[b]{0.275\textwidth}\begin{center}
\includegraphics[width=\textwidth]{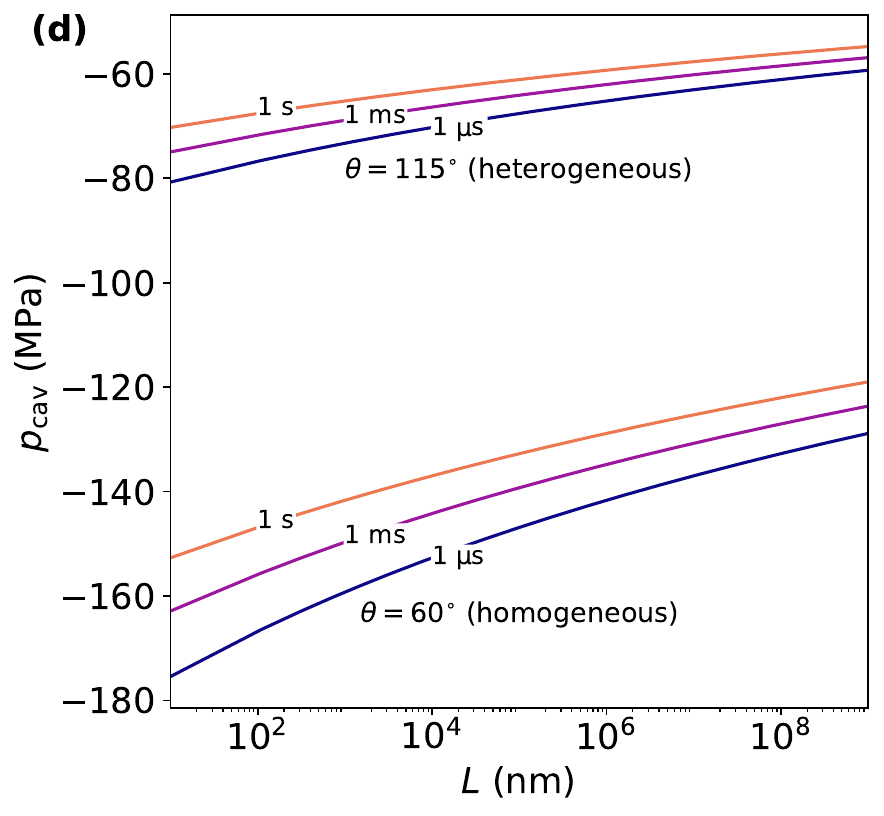}
\end{center}\end{minipage}
\caption{
(a) Sequential snapshots from a pressure-ramp simulation showing a homogeneous cavitation event in water near the hydrophilic surface (contact angle 60°), 
occurring when the pressure decreases down to $p_\cav^*=-$179 MPa at an applied a pressure rate of $-1$ MPa ns$^{-1}$.
(b) Sequential snapshots showing a heterogeneous cavitation event at the hydrophobic surface (contact angle 115°), occurring at $p_\cav^*=-$87 MPa when applying a pressure rate of $-$1 MPa ns$^{-1}$.
(c) Mean dynamic cavitation pressures obtained from pressure-ramp simulations for heterogeneous cavitation at the hydrophobic surface. The solid line represents the fit of the kinetic theory (\Eq~\ref{eq:p_star}), with $\kappa_\mathrm{s}$ as the fitting parameter.
(d) Cavitation pressure of a water-filled cubic container with 
hydrophilic and hydrophobic walls, plotted as a function of edge length $L$ for different waiting times.
}
\label{fig:smooth}
\end{center}\end{figure*}

The cavitation rate in \Eq~\ref{eq:k_comp} exhibits two limiting regimes.
As discussed in more detail elsewhere~\cite{loche2024water}, for hydrophilic surfaces with contact angles below a critical value of around 50 to 60°, the bulk cavitation rate dominates over the surface cavitation rate, $k_\w\gg k_\mathrm{s}$, and \Eq~\ref{eq:k_comp} simplifies to $k \approx k_\w$. Using \Eqs~\ref{eq:kw} and \ref{eq:dG_star_w}, and expressing the cavitation rate in terms of the mean cavitation time, $\tau=k^{-1}$, we derive a limiting expression for the cavitation pressure via the homogeneous pathway
\begin{equation}
    \label{eq:p_0_w}
    p_\mathrm{cav} =  - \sqrt{\frac{16 \pi  \gamma^3   }{3 \kB T\,  \ln (\kappa_\w L^3 \tau) }}
\end{equation}
which is a well-established result in CNT~\cite{caupin2006cavitation, herbert2006cavitation}. The cavitation pressure corresponds to the {\it kinetic stability limit}, where thermal fluctuations can overcome a finite free energy barrier within observation time $\tau$. 

Conversely, for larger contact angles, the cavitation rate at the surface prevails over the bulk rate ($k_\mathrm{s}\gg k_\mathrm{w}$), simplifying the total cavitation rate to $k \approx k_\mathrm{s}$.
Using \Eqs~\ref{eq:ks} and \ref{eq:free_eng_barr_het}, the limiting expression for cavitation pressure becomes~\cite{loche2024water}
\begin{equation}
    \label{eq:p_0_ws}
    p_\mathrm{cav} =  - \sqrt{\frac{16 \pi  \gamma^3  \, h(\theta) }{3 \kB T\,  \ln (\kappa_\mathrm{s} L^2 \tau) }}
\end{equation}

The attempt frequency densities, $\kappa_\mathrm{w}$ and  $\kappa_\mathrm{s}$, depend on molecular specifics and cannot be  estimated through continuum theories.
%Although some analytic estimates exist~\cite{blander1975bubble, pettersen1994experimental}, they are insufficient for our quantitative analysis.
To evaluate these quantities, we resort to computer simulations, employing the TIP4P/2005 water model along with a tunable self-assembled monolayer to present a hydrophilic ($\theta=60$°) and a hydrophobic ($\theta=115$°) surface.
In our recent study~\cite{sako2024impact}, we determined the bulk attempt frequency density of TIP4P/2005 water as $\kappa_\mathrm{w}= 1.25 \times 10^{18}$ s$^{-1}$nm$^{-3}$, using a pressure ramp method. This value will be reused here.
Applying the same method in the current study, we  determine $\kappa_\mathrm{s}$ for both hydrophilic and hydrophobic solid surfaces. 

In our pressure-ramp simulations with the hydrophilic surface, cavitation events consistently occur in the bulk water phase rather than at the surface, as shown by snapshots in \Fig~\ref{fig:smooth}a. This indicates that $k_\w\gg k_\mathrm{s}$ and $k_\mathrm{s}$ cannot be determined, but it is also irrelevant.

In contrast, on the hydrophobic surface, cavitation always occurs at the interface, as shown in the snapshots in \Fig~\ref{fig:smooth}b.
Figure~\ref{fig:smooth}c presents the results of the pressure-ramp simulations, plotting the inverse square of the dynamic cavitation pressures, $1/{p^*_\mathrm{cav}}^2$, against the negative pressure rate, $-\dot{p}$. 
%Each symbol represents the ensemble average of dynamic cavitation pressures over 6 to 8 independent simulation runs. 
The solid line represents the fit based on the kinetic theory (\Eq~\ref{eq:p_star}, where $G^*$ is given by
\Eq~\ref{eq:free_eng_barr_het}), yielding $\kappa_\mathrm{s} = 9.46 \times 10^{21}$ nm$^{-2}$s$^{-1}$. 
Notably, $\kappa_\mathrm{s}$ primarily influences the offset of the curve but not its slope.

With the parameters $\kappa_\w$ and $\kappa_\mathrm{s}$ at hand, we proceed to compute cavitation pressures. 
In \Fig~\ref{fig:smooth}d, we show the results for the box of water confined within atomistically smooth walls.
The pressures for homogeneous cavitation in the hydrophilic system (calculated using \Eq~\ref{eq:p_0_w}) range from $-120$ to $-170$ MPa, depending on the container size and observation time, consistent with previous theoretical results~\cite{fisher1948fracture, caupin2005liquid, caupin2006cavitation, caupin2013stability, azouzi2013coherent, sako2024impact}. 
In contrast, for hydrophobic walls ($\theta=115$°), heterogeneous cavitation occurs at pressures, calculated using \Eq~\ref{eq:p_0_ws}, that are approximately half those for homogeneous cavitation, between $-60$ and $-80$ MPa. These cavitation pressures are slightly more negative compared to those reported in our recent study because of a different water model employed~\cite{loche2024water}.
Since the cavitation pressure depends logarithmically on system volume and observation time, the narrow pressure range effectively represents the liquid's tensile strength---the maximum tension (i.e., negative pressure) the system can withstand before cavitation occurs.

These results demonstrate that sufficiently hydrophobic container surfaces can dramatically reduce the tensile strength. Yet, cavitation pressures remain considerably smaller (i.e., more negative) compared to those in typical experiments, which are $-30$ MPa or higher (i.e., lower in magnitude).
To address this discrepancy, we next explore how nanoscale surface defects that can host nanobubbles influence the stability of the system.

\subsection{Hydrophilic container with a single pit}

\begin{figure*}[h!]\begin{center}
\begin{minipage}[b]{0.225\textwidth}\begin{center}
\includegraphics[width=\textwidth]{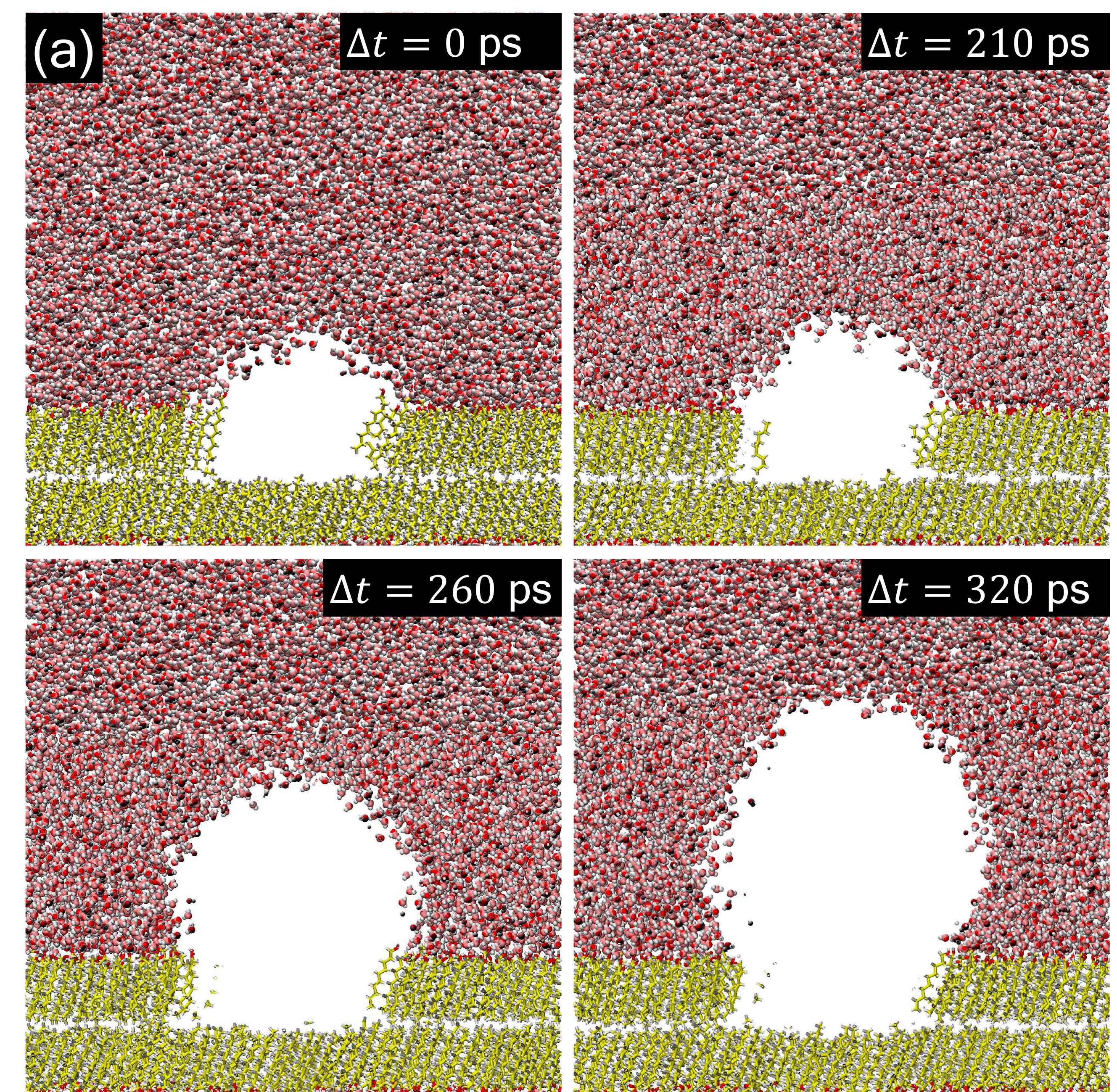}
\end{center}\end{minipage}
\begin{minipage}[b]{0.235\textwidth}\begin{center}
\includegraphics[width=\textwidth]{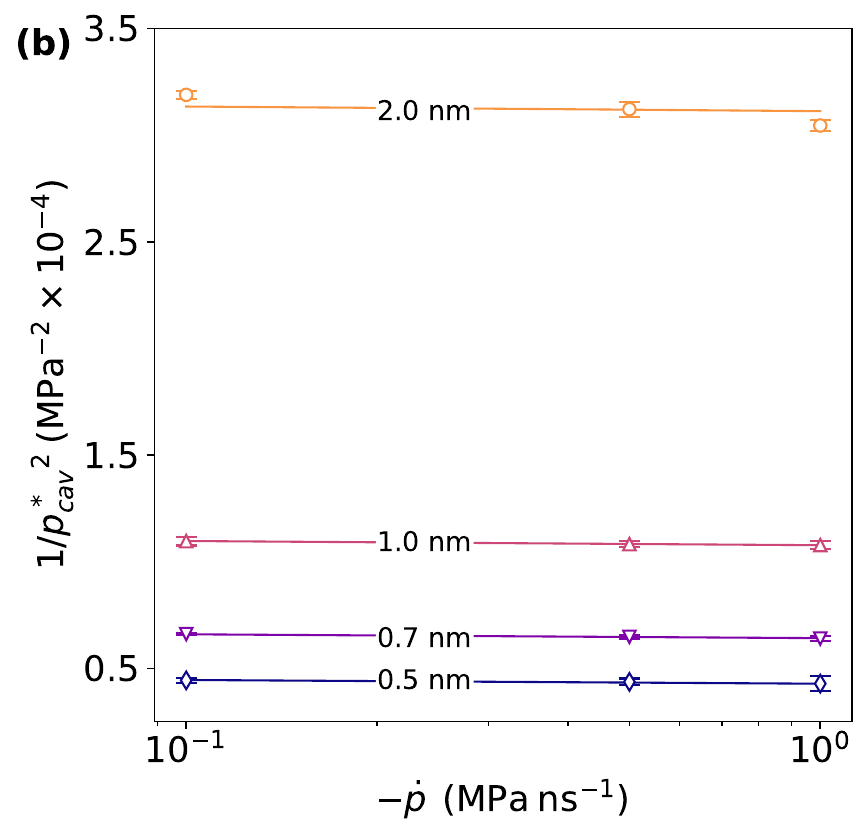}
\end{center}\end{minipage}
\begin{minipage}[b]{0.24\textwidth}\begin{center}
\includegraphics[width=\textwidth]{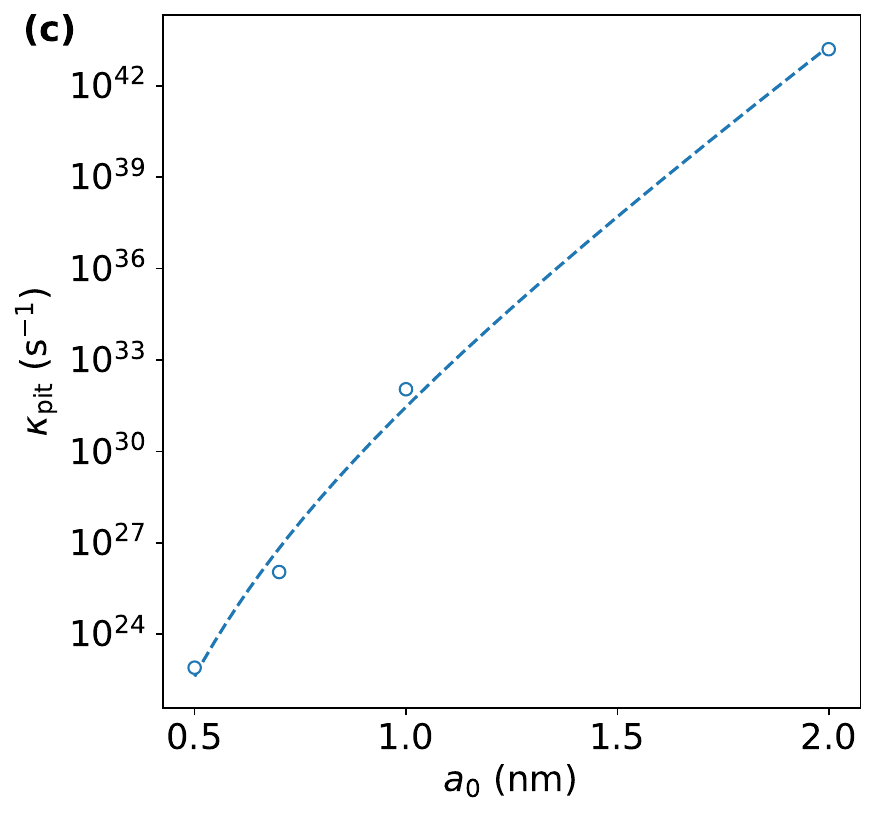}
\end{center}\end{minipage}
\begin{minipage}[b]{0.245\textwidth}\begin{center}
\includegraphics[width=\textwidth]{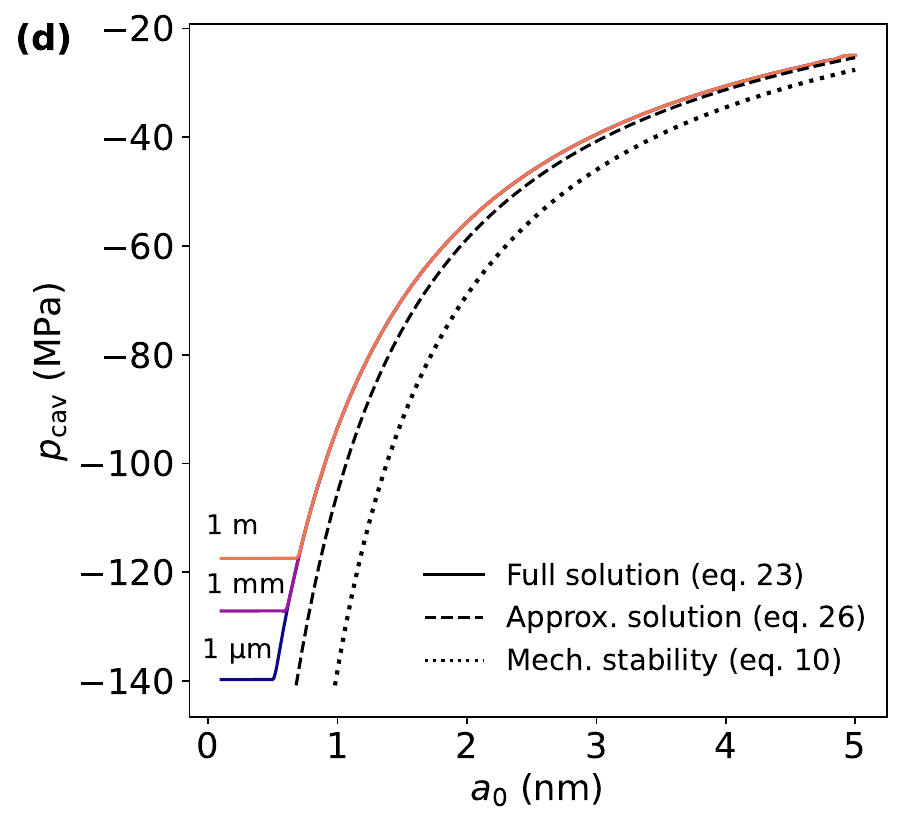}
\end{center}\end{minipage}
\caption{
Hydrophilic surface ($\theta=60\trm{°}$) with a pit:
(a) Sequential snapshots from a pressure-ramp simulation showing a cavitation event at a pit with a radius of $a_0 = 2$ nm. The initial snapshot shows depicts a metastable nanobubble  just before its spontaneous growth (shown in the subsequent three snapshots). The bubble spontaneously forms in the pit due to the hydrophpbic nature of the pit and remains metastable until pressure is ramped down to $p_\cav^*=-57$ MPa at a rate of $-0.5$ MPa ns$^{-1}$. At this pressure, the bubble expands further, leading to cavitation.
(b) Mean dynamic cavitation pressures from pressure-ramp simulations of a hydrophilic surface with a  pit of different radii (indicated by labels), plotted against the pressure rate (symbols). Solid lines represent fits based on the kinetic theory (\Eq~\ref{eq:p_star} with the free energy barrier given by $G^*_\trm{pin}$ in \Eq~\ref{eq:G_star}), with $\kappa_\mathrm{pit}$ as the fitting parameter.
(c) Cavitation attempt frequency $\kappa_\mathrm{pit}$ obtained from the fits (symbols) as a function of pit radius $a_0$. The dashed line is the fit of \Eq~\ref{eq:kappa_fit} to the data points.
(d) Cavitation pressure of a water-filled cubic container with hydrophilic walls containing a single hydrophobic pit, plotted as a function of pit radius for different container edge lengths $L$ (indicated by labels) and a waiting time of 1~s. The solid lines are the full numerical solution of \Eq~\ref{eq:k_tot}, the dashed line represents the approximate solution given by \Eq~\ref{eq:pcav_approx1}, while the dotted line represents the mechanical stability limit, given by $p=-2\gamma/a_0$ (\Eq~\ref{eq:p_0}).
}
\label{fig:defect_polar}
\end{center}\end{figure*}

To account for surface defects, we extend the cubic container model to include a single cylindrical pit of radius $a_0$ in the wall that hosts a surface nanobubble (as illustrated in \Fig~\ref{fig:geometries}c). 
The exact depth of the pit is irrelevant, provided it is sufficiently deep and hydrophobic to remain dewetted.
This pit introduces an additional pathway for cavitation, with the cavitation rate $k_\mathrm{pit}$.
The total cavitation rate reads~\cite{loche2024water}
\begin{equation}
    \label{eq:k_tot}
    k = k_\w+ k_\mathrm{s} + k_\mathrm{pit}
\end{equation}
While the first two cavitation rates remain the same as in the ideal, defect-free system discussed earlier (\Eq~\ref{eq:k_comp}), the cavitation rate due to the pit is given by
\begin{equation}
k_\mathrm{pit} = \kappa_\mathrm{pit}(a_0) \exp({-\beta G^*_\mathrm{pit}})
\label{eq:k_pit}
\end{equation}
Here, $G^*_\mathrm{pit}$ is the free energy barrier for a bubble nucleated at the pit, given by \Eq~\ref{eq:G_star_pit}, and the kinetic prefactor $\kappa_\mathrm{pit}(a_0)$ stands for the attempt frequency, which depends on the pit radius in general.

To simulate this cavitation pathway, we extend our simulation model by introducing a pit into the surface, created by removing decanol molecules within a specified radius from the center of one of the monolayers. We simulated circular pits with radii of 0.5, 0.7, 1, and 2 nm.
Snapshots from a pressure-ramp simulation for the system with
a pit with a radius of 2 nm are shown in \Fig~\ref{fig:defect_polar}a.
Because of the nonpolar nature of the decanol tails, an equilibrium vapor bubble forms inside the pit (as seen in the first snapshot of \Fig~\ref{fig:defect_polar}a), consistent with the configuration depicted in \Fig~\ref{fig:geometries}c. During cavitation, the bubble expands in a pinned manner, maintaining its base radius fixed without spreading across the surface, as illustrated in \Fig~\ref{fig:geometries}d.

The dynamic cavitation pressures versus pressure rates obtained from pressure-ramp simulations for all four pit sizes are shown in \Fig~\ref{fig:defect_polar}b as symbols. The fits of the kinetic theory (\Eq~\ref{eq:p_star}) are represented by solid lines, which yield the values of $\kappa_\mathrm{pit}$ for each pit size.
The fitted values of $\kappa_\mathrm{pit}$ against pit radius are displayed in \Fig~\ref{fig:defect_polar}c. We observe that $\kappa_\mathrm{pit}$ depends considerably on the pit size. As we do not have a theoretical model to describe this relationship, we employ an empirical fitting function of the form
\begin{equation}
\label{eq:kappa_fit}
\kappa_\mathrm{pit}(a_0) = \kappa_{\mathrm{pit},0} \exp{\left( c_{-1}/a_0 + c_1 a_0  \right)}   
\end{equation}
where $\kappa_{\mathrm{pit},0}$, $c_{-1}$, and $c_1$ are the fitting parameters, listed in \Tab~\ref{tab:kappa_fit}. 
The fit of this function is depicted as a dashed line in \Fig~\ref{fig:defect_polar}c and matches the MD data very well.
For more details about the fitting procedure and the values of the fitted coefficients.

\begin{table}
\footnotesize
    \begin{tabular}{ c  l  l } 
        Fitting parameter & Hydrophilic surface & Hydrophobic surface \\ 
        \hline
        \\
        $\kappa_0$ & $3.14 \times 10^{25}$ s$^{-1}$ & $2.11 \times 10^{26}$ s$^{-1}$\\
        $c_1$ & $22.74$ nm$^{-1}$ & $3.63$ nm$^{-1}$ \\ 
        $c_{-1}$ & $-9.00$ nm & $-5.37$ nm \\ 
    \end{tabular}
    \caption{Fitting parameters obtained by fitting \Eq~\ref{eq:kappa_fit} to the $\kappa_\trm{pit}$ data in \Figs~\ref{fig:defect_polar}c (hydrophilic surface) and ~\ref{fig:defect_nonpolar}c (hydrophobic surface).}
    \label{tab:kappa_fit}
\end{table}

We will not delve into the physical interpretation of the empirical form of \Eq~\ref{eq:kappa_fit}. 
However, it is noteworthy that the linear term ($c_1 a_0$) in the exponent is analogous to the linear dependence of the free energy barrier on $a_0$. Thus, its necessity in the fit to describe the data at larger pit sizes may stem from line-tension effects,
which scale with the pit's perimeter, as well as curvature corrections to the water surface tension. 
%Note also that quadratic or higher-order terms in the exponent of \Eq~\ref{eq:kappa_fit} are not appropriate, as they would conflict with the leading-order dependence of the free energy on $a_0$.

We should emphasize that, although the surface is only mildly hydrophilic, we consistently observe cavitation occurring as a pinned bubble in all the cases. This is because the pressure remains below the threshold required to transition to the spreading mode; that is, $p/p_0<\sin\theta$. 
In this regime, the cavitation process is not directly influenced by the contact angle. 

We now incorporate the empirical function $\kappa_\mathrm{pit}(a_0)$ 
(\Eq~\ref{eq:kappa_fit}) into our CNT model of a cubic container (\Eqs~\ref{eq:k_tot} and \ref{eq:k_pit}).
For a sufficiently hydrophilic surface (where $k_\mathrm{s}\ll k_\mathrm{w}$), nucleation at the pit competes with homogeneous nucleation, such that the total cavitation rate (\Eq~\ref{eq:k_tot}) simplifies to $k = k_\w+ k_\mathrm{pit}$.
This transcendental equation must be solved numerically to determine the cavitation pressures across the full range of pit sizes.

The numerical results for cavitation pressure as a function of pit radius is shown in \Fig~\ref{fig:defect_polar}d as solid lines.
For pit radii smaller than a fraction of a nanometer, the system undergoes homogeneous cavitation, with the cavitation pressure determined by \Eq~\ref{eq:p_0_w}.
However, as the pit radius exceeds around 0.5 nm---referred to as the {\it critical pit radius}, $a_0^\trm{min}$---the pit begins to act as a cavitation nucleus, leading to a sharp decrease in cavitation pressure with increasing pit size. 
Notably, variations in box size have a minimal effect on $a_0^\trm{min}$.

An approximate analytical solution for the cavitation pressure can be derived in the limit of a small free energy barrier, which occurs for pressures close to the mechanical stability limit (i.e., $p\approx p_0$).
Expanding $G^*_\textrm{pit}$ for pressures around $p_0$ enables an analytical solution of the rate equation (\Eq~\ref{eq:k_pit}), yielding the following approximate expression for the cavitation pressure
\begin{equation}
    p_\trm{cav}\approx-\frac{2\gamma}{a_0}\left[1-\left(\frac{\xi_\trm{p}}{a_0}\right)^{2/3}\right]
    \label{eq:pcav_approx1}
\end{equation}
Here, we introduced the correction length for the pinned mode 
\begin{equation}
    \xi_\trm{p}=\frac{3 k_\trm{B} T c_1}{8 \sqrt{2}  \pi\gamma }
\end{equation}
which amounts to $\xi_\trm{p}\approx 0.11$ nm. 
This expression shows that the cavitation pressure asymptotically approaches the mechanical stability limit ($p_\mathrm{cav} \approx p_0$) for sufficiently large defect sizes ($ a_0 \gg \xi_\mathrm{p}$).
The result of the analytical prediction is shown as a dashed line in \Fig~\ref{fig:defect_polar}d and agrees closely with the full numerical solution (solid lines). 

For pit radii $a_0\gg\xi_\trm{p}$, \Eq~\ref{eq:pcav_approx1} simplifies to $p_\trm{cav}\approx-2\gamma/a_0=p_0$, corresponding to the mechanical stability limit, where the free energy barrier vanishes ($G^*_\mathrm{pit}=0$).
The mechanical stability limit is plotted as a dotted line \Fig~\ref{fig:defect_polar}d and represents the asymptotic expression for cavitation pressure for large pit radii.

Thus, the correction term in the brackets of \Eq~\ref{eq:pcav_approx1} captures the effect of free energy barrier crossing, with the parameter $\xi_\trm{p}$ defining the defect length scale at which this effect becomes significant. 
For pit radii much larger than $\xi_\mathrm{p}$, the difference between the kinetic and mechanical stability limits becomes negligible, allowing the cavitation pressure to be approximated by the mechanical stability limit. However, for bubbles only a few nanometers in size, this difference is quantitatively significant, and the stochastic nature of free-energy barrier crossing must be taken into account.

A relevant question in this context concerns the minimal pit radius that is capable of acting as a nucleation site. 
A simple mechanical stability criterion suggests a minimum radius of $a_{0}^\trm{min}=-2\gamma/p_\trm{cav}^0\approx 1$ nm, based on a cavitation pressure of $p_\trm{cav}^0\approx-120$ MPa for homogeneous nucleation. However, the full solution derived from CNT yields $a_{0}^\trm{min}\approx 0.5$ nm, as seen from \Fig~\ref{fig:defect_polar}d (solid lines).
Thus, the prediction from mechanical stability is a factor of two too large.
This result demonstrates the limitations of applying mechanical stability concepts at nanometer scales, showing that it is necessary to consider kinetic contributions to nucleation processes at these length scales.

\subsection{Hydrophobic container with a single pit}

\begin{figure*}[h!]\begin{center}
\begin{minipage}[b]{0.225\textwidth}\begin{center}
\includegraphics[width=\textwidth]{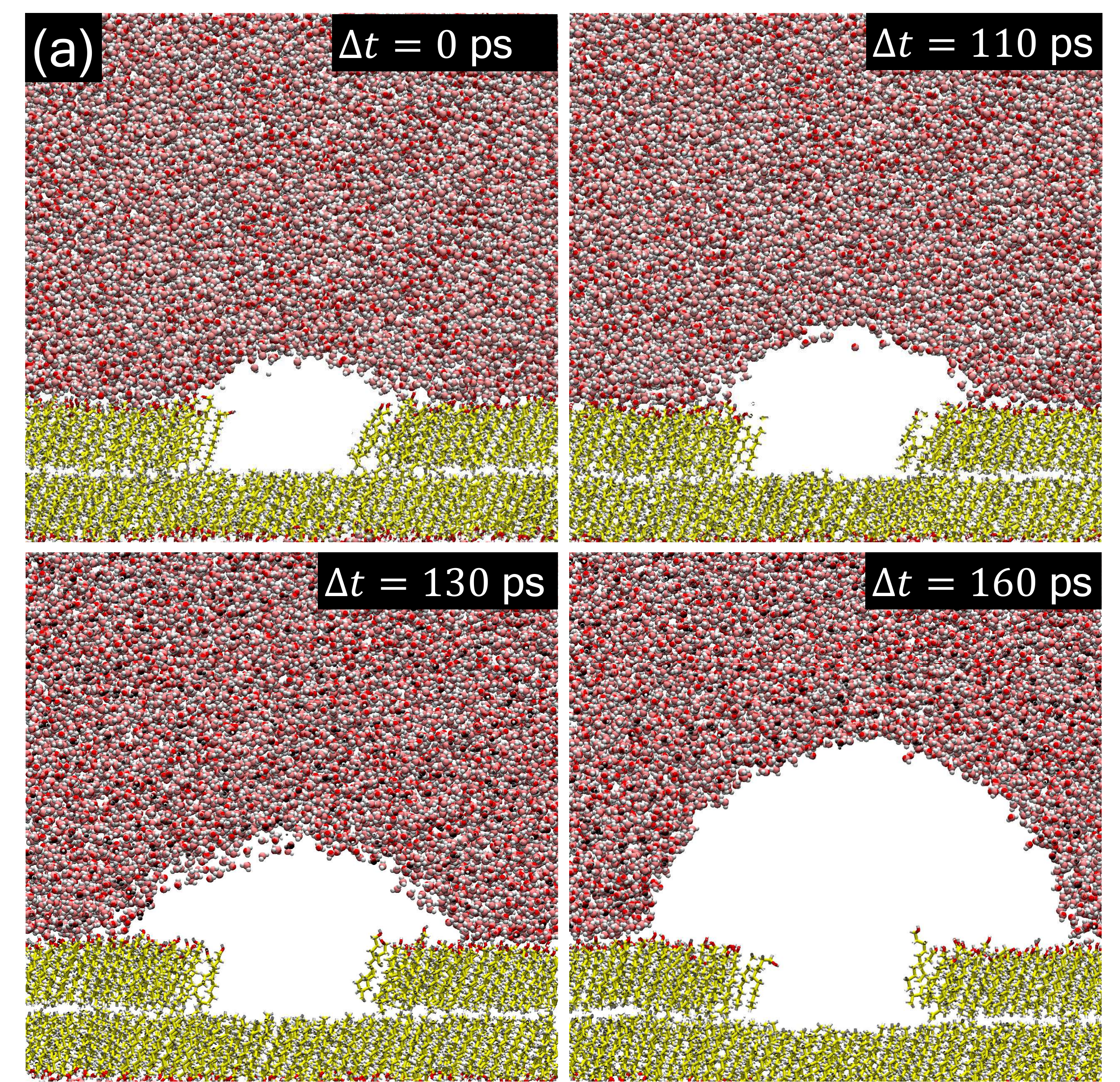}
\end{center}\end{minipage}
\begin{minipage}[b]{0.23\textwidth}\begin{center}
\includegraphics[width=\textwidth]{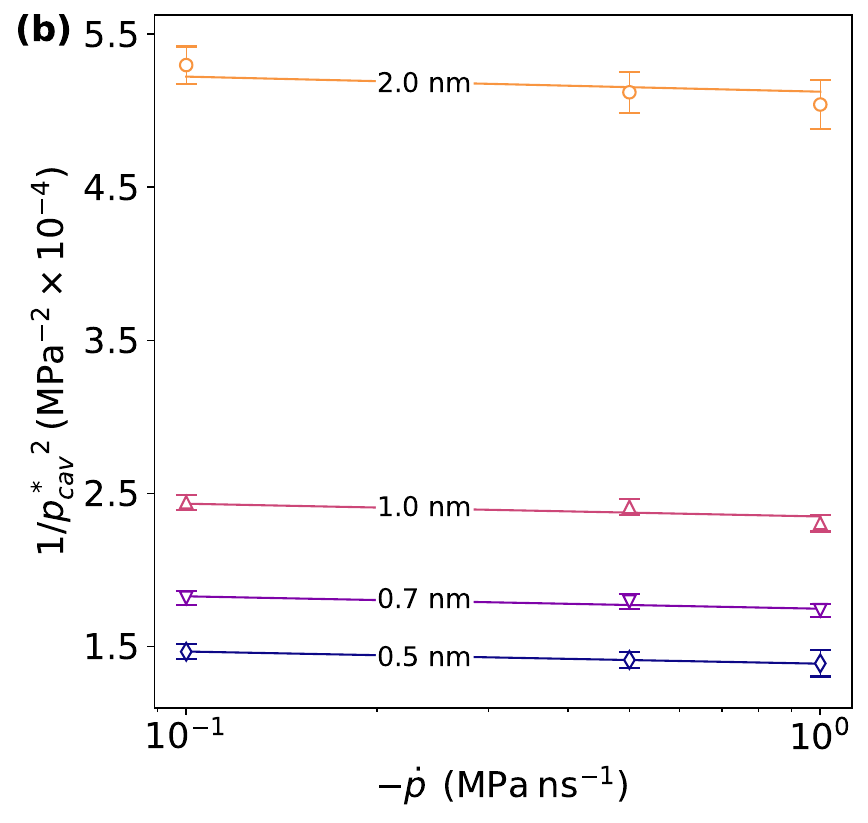}
\end{center}\end{minipage}
\begin{minipage}[b]{0.24\textwidth}\begin{center}
\includegraphics[width=\textwidth]{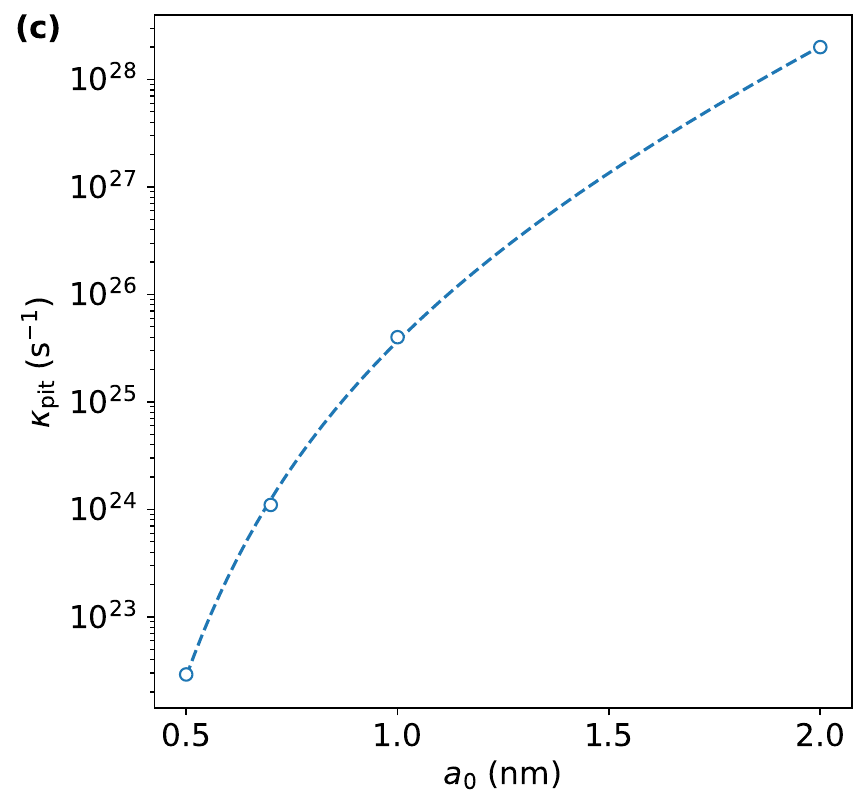}
\end{center}\end{minipage}
\begin{minipage}[b]{0.25\textwidth}\begin{center}
\includegraphics[width=\textwidth]{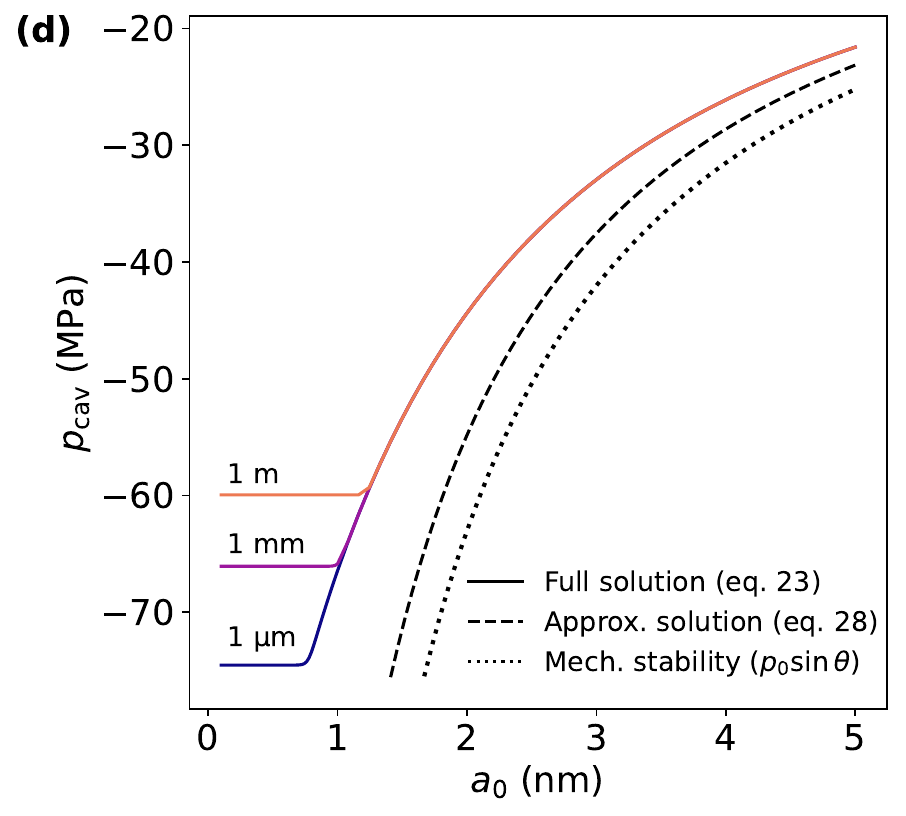}
\end{center}\end{minipage}
\caption{
Hydrophobic surface ($\theta=115\trm{°}$) with a pit:
(a) Sequential snapshots from a pressure-ramp simulation showing a cavitation event at a pit of a radius of $a_0 = 2$ nm. The initial shows depicts a metastable nanobubble that has spontaneously formed within the pit. The cavitation occurred as the dynamic pressure dropped down to $p_\cav^*=-45$ MPa under a pressure rate of $-1$ MPa ns$^{-1}$.
(b) Mean dynamic cavitation pressures from pressure-ramp simulations of a hydrophilic surface with a  pit of different radii (indicated by labels), plotted against the pressure rate (symbols). Solid lines represent fits based on the kinetic theory (\Eq~\ref{eq:p_star} with the free energy barrier given by $G^*_\trm{spr}$ in \Eq~\ref{eq:free_eng_barr_spr}), with $\kappa_\mathrm{pit}$ as the fitting parameter.
(c) Cavitation attempt frequency $\kappa_\mathrm{pit}$ obtained from the fits (symbols) as a function of pit radius $a_0$. The dashed line is the fit of \Eq~\ref{eq:kappa_fit} to the data points.
(d) Cavitation pressure of a water-filled cubic container with hydrophobic walls containing a single hydrophobic pit, plotted as a function of pit radius for different container edge lengths $L$ (indicated by labels) and a waiting time of 1~s. The solid lines are the full numerical solution of \Eq~\ref{eq:k_tot}, the dashed line represents the approximate solution given by \Eq~\ref{eq:pcav_approx_s}, while the dotted line represents the mechanical stability limit, given by $p=-(2\gamma/a_0)\sin\theta$.
}
\label{fig:defect_nonpolar}
\end{center}\end{figure*}

Next, we examine a water-filled container with a single pit, as before, but now with hydrophobic walls (with a contact angle of $\theta=115$°). This setup introduces two key differences compared to the case with hydrophilic walls.
First, as previously noted, the cavitation rate for heterogeneous cavitation at the smooth (i.e., defect-free) surface dominates over that of homogeneous cavitation ($k_\mathrm{w}\ll k_\mathrm{s}$), simplifying the total cavitation rate in \Eq~\ref{eq:k_tot} to $k = k_\mathrm{s}+ k_\mathrm{pit}$. In this scenario, cavitation at the pit competes with heterogeneous cavitation at the smooth part of the surface.
Second, unlike in the hydrophilic case, the vapor bubble unpins and spreads over the surface before reaching criticality.

Simulation snapshots from a cavitation event on the hydrophobic surface are shown in \Fig~\ref{fig:defect_nonpolar}a, illustrating how the preexisting vapor bubble spreads across the surface as a spherical cap, while maintaining approximately a constant contact angle. 
The dynamic cavitation pressures from pressure-ramp simulations are presented in \Fig~\ref{fig:defect_nonpolar}b as symbols. The fits of the kinetic theory (\Eq~\ref{eq:p_star}) using $G^*=G_\mathrm{spr}$ (\Eq~\ref{eq:free_eng_barr_spr}) are shown as solid lines. The corresponding fitted values of $\kappa_\mathrm{pit}$ are presented in \Fig~\ref{fig:defect_nonpolar}c. 
Similarly to the hydrophilic case, we use the empirical fitting function given by \Eq~\ref{eq:kappa_fit} to fit these data points, shown by the solid line (the corresponding parameters are listed in \Tab~\ref{tab:kappa_fit}).

The obtained empirical function for $\kappa_\mathrm{pit}$ allows us to numerically calculate the cavitation pressure as a function of pit radius, as presented in \Fig~\ref{fig:defect_nonpolar}d. The outcome is qualitatively similar to that observed for hydrophilic walls: when the pit radius is smaller than approximately 1 nm, the pit has no effect on cavitation, and the system cavitates heterogeneously on the smooth part of the wall. For pit radii above the critical size of $a_{0}^\trm{min}\approx 1$ nm, the vapor bubble in the pit acts as a nucleation site, causing the cavitation pressure to rapidly shift to less negative values.

 For small free energy barriers, which occur for pressures near the mechanical stability limit, we derive an analytical expression for $p_\trm{cav}$. The resulting expression for the cavitation pressure reads
\begin{equation}
    p_\trm{cav}\approx-\frac{2\gamma}{a_0}\sin\theta\left[1-\left(\frac{\xi_\trm{s}}{a_0}\right)^{1/2}\right]
    \label{eq:pcav_approx_s}
\end{equation}
where the correction length for the spreading mode is given by
\begin{equation}
    \xi_\trm{s}=-\frac{k_\trm{B}T c_1 \cos \theta}{\pi \gamma \sin^2 \theta} 
    \label{eq:ksi_s}
\end{equation}
and amounts to $\xi_\trm{p}\approx 0.035$ nm. Similar to the pinned case, the cavitation pressure asymptotically approaches the mechanical stability limit ($p_\trm{cav} \approx p_0\sin\theta$) for sufficiently large defect sizes.
This analytical prediction is shown in \Fig~\ref{fig:defect_nonpolar}d as a dashed line, showing a slower convergence to the full numerical solution (solid lines) compared to the hydrophilic case in \Fig~\ref{fig:defect_polar}d.
 The slower convergence arises primarily from the lower power-law scaling in the correction term, $1/2$ here versus $2/3$ for the pinned case on hydrophilic walls. Thus, higher-order terms are needed for better agreement.

The leading-order approximation of \Eq~\ref{eq:pcav_approx_s} reads $p_\trm{cav}\approx-(2\gamma/a_0) \sin\theta$, which corresponds to the mechanical stability limit where the free energy barrier vanishes ($ G^*_\mathrm{spr} = 0$). This mechanical stability limit is shown as a dotted line in \Fig~\ref{fig:defect_nonpolar}d and represents the asymptotic expression for the cavitation pressure at large pit radii.

The mechanical stability criterion predicts a critical pit radius of $a_{0}^\trm{min}=-2\gamma/p_\trm{cav}^0\sin\theta\approx 2$ nm, based on a cavitation pressure of $p_\trm{cav}^0\approx-60$ MPa for heterogeneous cavitation at the smooth hydrophobic surface. However, as shown in \Fig~\ref{fig:defect_nonpolar}d, the full CNT solution (solid lines) indicates a smaller value of $a_{0}^\trm{min}\approx 1$ nm. 
This demonstrates that the mechanical stability estimate is again overestimated by a factor of two, reinforcing the conclusion that significantly smaller pits can serve as cavitation nuclei also on hydrophobic surfaces, owing to the influence of the free-energy barrier crossing.

\subsection{Container with multiple pits}

Until now, our model has been limited to surfaces containing a single pit. However, realistic surfaces are typically far more complex, often featuring multiple topographical defects and cavities that could act as nucleation sites~\cite{de2013surface, siretanu2016atomic}. To better reflect this complexity, we extend our CNT model to include surface walls with $N$ identical pits, each potentially influencing the overall cavitation process. The total cavitation rate for such a system is similar to \Eq~\ref{eq:k_tot} for one pit but with the cavitation rate for each pit multiplied by $N$, reading~\cite{loche2024water}
\begin{equation}
    \label{eq:k_tot_N_pits}
    k = k_\w + k_\mathrm{s} + N k_\mathrm{pit}
\end{equation}
We can express the areal pit density (number of pits per unit area) as $\rho_A=N/(6L^2)$.

\begin{figure}
    \begin{subfigure}[b]{0.235\textwidth}
    \includegraphics[width=\textwidth]{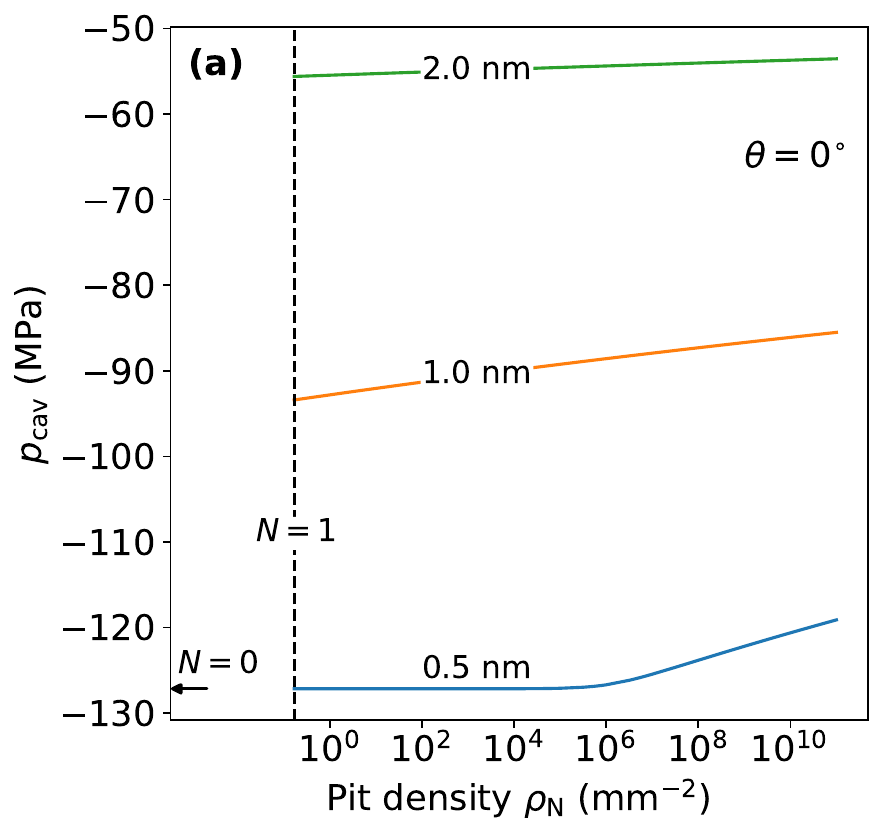}
    % \caption{}
    \label{fig:defect_density_polar}
    \end{subfigure}
  \hspace{0.0cm}
    \begin{subfigure}[b]{0.218\textwidth}
    \includegraphics[width=\textwidth]{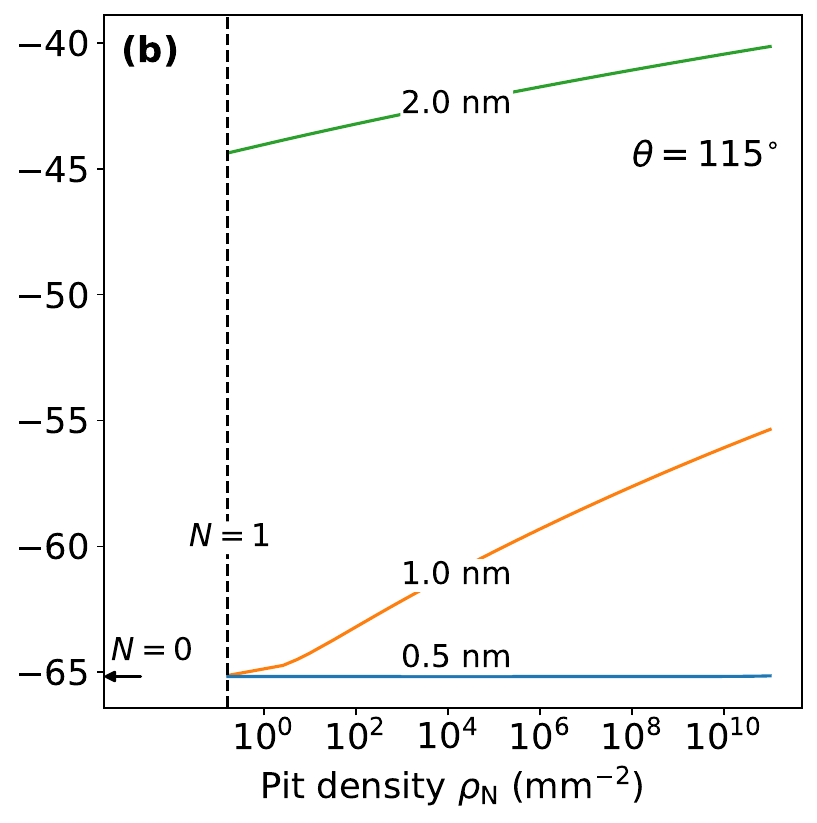}
    % \caption{}
    \label{fig:defect_density_nonpolar}
    \end{subfigure}
\caption{
Cavitation pressure in a cubic water-filled container with (a) hydrophilic ($\theta=0\trm{°}$) and (b) hydrophobic ($\theta=115\trm{°}$) walls, plotted as a function of areal pit density of various pit radii, predicted from \Eq~\ref{eq:k_tot_N_pits}. The container has a side length of $L=1$~mm, and the observation time is 1 s.
The vertical dashed line indicates the limiting pit density corresponding to a single pit in the system ($N=1$). The arrow indicates the cavitation pressure of the system without any pits.
}
\label{fig:defect_density}
\end{figure}

By numerically solving \Eq~\ref{eq:k_tot_N_pits}, we present in \Fig~\ref{fig:defect_density} the cavitation pressure of a cubic water-filled container with a side length of $L=1$~mm for a waiting time of 1 s, plotted a function of areal pit density. Panel (a) considers hydrophilic walls ($\theta=0^\circ$), while panel (b) hydrophobic walls ($\theta=115^\circ$).
The arrows on the left side of each graph indicate the cavitation pressures in defect-free systems ($N=0$). 
The vertical dashed lines mark the area density corresponding to a single pit in the system ($N=1$), calculated as $(6L^2)^{-1}\approx0.17$~mm$^{-2}$.  Solid lines represent the cavitation pressures as the pit density increases further ($N\ge1$), with results shown for three different pit radii, as indicated by the labels.

Subcritical pits, $a_0 < a_0^\trm{min}$ (where $a_0^\trm{min}\approx0.5$ and $1$ nm for the hydrophilic and hydrophobic surface, respectively), have little to no effect on the cavitation pressure. For the pits with radii $a_0 = 0.5$ nm on hydrophilic surfaces (\Fig~\ref{fig:defect_density}a), a subtle but sudden increase in cavitation pressure is observed starting at a pit density of $10^6$ nm$^{-2}$. At this density, the cavitation rate at the pits matches the rate of homogeneous cavitation, $N k_\mathrm{pit}\approx k_\w$. Similarly, for the pit radii of 1 nm on hydrophobic surfaces, this transition occurs at densities above 1 nm$^{-2}$.

In contrast, introducing a single supercritical pit ($a_0 > a_0^\mathrm{min}$) causes a dramatic jump in cavitation pressure compared to defect-free systems, consistent with the findings in \Figs~\ref{fig:defect_polar}d and \ref{fig:defect_nonpolar}d.
However, adding more such pits of the same size has only a minimal additional impact on the overall cavitation pressure, regardless of surface hydrophobicity. Even as the pit density increases by several orders of magnitude, the change in cavitation pressure remains modest---typically on the order of a few MPa.

The observed minimal impact of the number of pits can be readily explained.
In the regime where pits dominate cavitation, the rate in \Eq~\ref{eq:k_tot_N_pits} simplifies to $k/N \approx k_{\text{pit}}$, indicating that cavitation pressure depends on $N$ in the same weak manner as it does on observation time $\tau=k^{-1}$.  Thus, the critical factor for cavitation is the size of the largest pit rather than the number of pits. Larger pit sizes further diminish the influence of pit density on cavitation pressure.
This finding aligns with experimental results by Borkent et al.~\cite{borkent2009nucleation}, who explored cavitation on surfaces with mesoscopically large pits (radii above 50 nm). They demonstrated that the nucleation threshold was dictated by the largest cavities, with no discernible dependence on pit density. Indeed, in the mesoscopic regime, the cavitation pressure approaches the mechanical stability limit (given by \Eq~\ref{eq:p_0}), eventually becoming independent of the number of pits.

Thus, simply increasing pit density without altering other factors (e.g., pit size or surface chemistry) has a limited effect on cavitation pressure. Conversely, reducing the density of topographical defects is less important than eliminating the largest defects. While many small defects have a relatively modest effect on cavitation, a single larger defect represents the ``weakest spot,'' which dominates and overshadows the others. Consequently, this feature highlights the challenge of engineering surfaces capable of sustaining high cavitation pressures as already a single large defect significantly lowers the system's overall stability under negative pressure.
Note that because hydrophobic defects can spontaneously form vapor bubbles, removing dissolved gases from the system (degassing) has no significant influence on the cavitation behavior associated with these tiny defects.

\section{Conclusions}
In this study, we investigated how nanoscale surface defects, specifically hydrophobic pits, influence cavitation behavior in water under negative pressures. Through MD simulations and CNT, we showed that pure water contained within a hydrophilic, defect-free container exhibits exceptional resistance to cavitation, withstanding extreme negative pressures. While sufficiently hydrophobic yet atomistically flat container surfaces trigger heterogeneous cavitation, they reduce the magnitude of negative pressure required for cavitation. However, the resulting pressure magnitudes remain far below those seen in typical experiments.

The propensity of cavitation changes drastically when a single nanoscopic hydrophobic pit is introduced on the surface. A hydrophobic pit spontaneously dewets and forms a vapor bubble inside it.
On hydrophilic surfaces, cavitation proceeds via pinned bubble expansion at the pit, whereas hydrophobic surfaces promote bubble unpinning and spreading, leading to distinct free-energy barriers and cavitation thresholds.
Notably, the critical pit size required to nucleate cavitation is around two times smaller than predicted by mechanical stability models, emphasizing the stochastic nature of free-energy barrier crossing at the nanoscale.
The presence of a vapor bubble dramatically reduces the tensile strength of the system, with the extent of the reduction being strongly dependent on the defect size.
Our results highlight that cavitation is dictated by the largest defect rather than the overall defect density, underscoring the importance of eliminating larger imperfections to prevent cavitation.

As a final remark, we note that real-world scenarios may involve various kinds of impurities, such as nonpolar or amphiphilic molecules~\cite{roger2012hydrophobic, maali2017viscoelastic, uematsu2019impurity, zhang2021critical}. These molecules tend to adsorb onto hydrophobic surfaces~\cite{10.1093/pnasnexus/pgad190}, including hydrophobic defects, rendering them hydrophilic and thereby inhibiting spontaneous vapor bubble formation. 
Thus, it seems plausible that surfactants may provide a natural strategy for stabilizing liquids against cavitation.
For example, plants appear to use water-insoluble surfactants, such as lipids, in their sap as a protection against cavitation under negative pressures~\cite{schenk2015nanobubbles}.
Investigating the stability of surfactant-coated defects and their ability to suppress cavitation is a compelling direction for future research, which we plan to pursue in upcoming studies.

%The  bulk--surface cavitation crossover contact angle $\theta^*$ is close to the `Berg limit'~\cite{vogler1998structure, rosenhahn2010role, kanduc2024understanding}: Surfaces with larger contact angles are prone to macromolecule adhesion, which tends to make them more hydrophilic~\cite{sako2023conditions} and thus can prevent surface cavitation. Such a mechanism might be important for preventing embolies in the sap of tall trees. 

\section*{Acknowledgments}
M.\v{S}., F.S., and M.K.\ acknowledge financial support from the Slovenian Research and Innovation Agency ARIS (contracts P1-0055 and J1-4382). 

\normalsize

\footnotesize
\setlength{\bibsep}{0pt}
\bibliography{bibliography/sample}

\providecommand{\latin}[1]{#1}
\makeatletter
\providecommand{\doi}
  {\begingroup\let\do\@makeother\dospecials
  \catcode`\{=1 \catcode`\}=2 \doi@aux}
\providecommand{\doi@aux}[1]{\endgroup\texttt{#1}}
\makeatother
\providecommand*\mcitethebibliography{\thebibliography}
\csname @ifundefined\endcsname{endmcitethebibliography}
  {\let\endmcitethebibliography\endthebibliography}{}
\begin{mcitethebibliography}{57}
\providecommand*\natexlab[1]{#1}
\providecommand*\mciteSetBstSublistMode[1]{}
\providecommand*\mciteSetBstMaxWidthForm[2]{}
\providecommand*\mciteBstWouldAddEndPuncttrue
  {\def\EndOfBibitem{\unskip.}}
\providecommand*\mciteBstWouldAddEndPunctfalse
  {\let\EndOfBibitem\relax}
\providecommand*\mciteSetBstMidEndSepPunct[3]{}
\providecommand*\mciteSetBstSublistLabelBeginEnd[3]{}
\providecommand*\EndOfBibitem{}
\mciteSetBstSublistMode{f}
\mciteSetBstMaxWidthForm{subitem}{(\alph{mcitesubitemcount})}
\mciteSetBstSublistLabelBeginEnd
  {\mcitemaxwidthsubitemform\space}
  {\relax}
  {\relax}

\bibitem[Caupin and Herbert(2006)Caupin, and Herbert]{caupin2006cavitation}
Caupin,~F.; Herbert,~E. Cavitation in water: a review. \emph{Comptes Rendus
  Physique} \textbf{2006}, \emph{7}, 1000--1017\relax
\mciteBstWouldAddEndPuncttrue
\mciteSetBstMidEndSepPunct{\mcitedefaultmidpunct}
{\mcitedefaultendpunct}{\mcitedefaultseppunct}\relax
\EndOfBibitem
\bibitem[Stroock \latin{et~al.}(2014)Stroock, Pagay, Zwieniecki, and
  Michele~Holbrook]{stroock2014physicochemical}
Stroock,~A.~D.; Pagay,~V.~V.; Zwieniecki,~M.~A.; Michele~Holbrook,~N. The
  physicochemical hydrodynamics of vascular plants. \emph{Annu. Rev. Fluid
  Mech.} \textbf{2014}, \emph{46}, 615--642\relax
\mciteBstWouldAddEndPuncttrue
\mciteSetBstMidEndSepPunct{\mcitedefaultmidpunct}
{\mcitedefaultendpunct}{\mcitedefaultseppunct}\relax
\EndOfBibitem
\bibitem[Wheeler and Stroock(2008)Wheeler, and
  Stroock]{wheeler2008transpiration}
Wheeler,~T.~D.; Stroock,~A.~D. The transpiration of water at negative pressures
  in a synthetic tree. \emph{Nature} \textbf{2008}, \emph{455}, 208--212\relax
\mciteBstWouldAddEndPuncttrue
\mciteSetBstMidEndSepPunct{\mcitedefaultmidpunct}
{\mcitedefaultendpunct}{\mcitedefaultseppunct}\relax
\EndOfBibitem
\bibitem[Pagay \latin{et~al.}(2014)Pagay, Santiago, Sessoms, Huber, Vincent,
  Pharkya, Corso, Lakso, and Stroock]{pagay2014microtensiometer}
Pagay,~V.; Santiago,~M.; Sessoms,~D.~A.; Huber,~E.~J.; Vincent,~O.;
  Pharkya,~A.; Corso,~T.~N.; Lakso,~A.~N.; Stroock,~A.~D. A microtensiometer
  capable of measuring water potentials below -10 MPa. \emph{Lab Chip}
  \textbf{2014}, \emph{14}, 2806--2817\relax
\mciteBstWouldAddEndPuncttrue
\mciteSetBstMidEndSepPunct{\mcitedefaultmidpunct}
{\mcitedefaultendpunct}{\mcitedefaultseppunct}\relax
\EndOfBibitem
\bibitem[Vincent \latin{et~al.}(2014)Vincent, Sessoms, Huber, Guioth, and
  Stroock]{vincent2014drying}
Vincent,~O.; Sessoms,~D.~A.; Huber,~E.~J.; Guioth,~J.; Stroock,~A.~D. Drying by
  cavitation and poroelastic relaxations in porous media with macroscopic pores
  connected by nanoscale throats. \emph{Phys. Rev. Lett.} \textbf{2014},
  \emph{113}, 134501\relax
\mciteBstWouldAddEndPuncttrue
\mciteSetBstMidEndSepPunct{\mcitedefaultmidpunct}
{\mcitedefaultendpunct}{\mcitedefaultseppunct}\relax
\EndOfBibitem
\bibitem[Dular \latin{et~al.}(2004)Dular, Bachert, Stoffel, and
  {\v{S}}irok]{dular2004relationship}
Dular,~M.; Bachert,~B.; Stoffel,~B.; {\v{S}}irok,~B. Relationship between
  cavitation structures and cavitation damage. \emph{Wear} \textbf{2004},
  \emph{257}, 1176--1184\relax
\mciteBstWouldAddEndPuncttrue
\mciteSetBstMidEndSepPunct{\mcitedefaultmidpunct}
{\mcitedefaultendpunct}{\mcitedefaultseppunct}\relax
\EndOfBibitem
\bibitem[Reuter \latin{et~al.}(2022)Reuter, Deiter, and
  Ohl]{reuter2022cavitation}
Reuter,~F.; Deiter,~C.; Ohl,~C.-D. Cavitation erosion by shockwave
  self-focusing of a single bubble. \emph{Ultrason Sonochem} \textbf{2022},
  \emph{90}, 106131\relax
\mciteBstWouldAddEndPuncttrue
\mciteSetBstMidEndSepPunct{\mcitedefaultmidpunct}
{\mcitedefaultendpunct}{\mcitedefaultseppunct}\relax
\EndOfBibitem
\bibitem[Adhikari \latin{et~al.}(2015)Adhikari, Goliaei, and
  Berkowitz]{adhikari2015mechanism}
Adhikari,~U.; Goliaei,~A.; Berkowitz,~M.~L. Mechanism of membrane poration by
  shock wave induced nanobubble collapse: A molecular dynamics study. \emph{J.
  Phys. Chem. B} \textbf{2015}, \emph{119}, 6225--6234\relax
\mciteBstWouldAddEndPuncttrue
\mciteSetBstMidEndSepPunct{\mcitedefaultmidpunct}
{\mcitedefaultendpunct}{\mcitedefaultseppunct}\relax
\EndOfBibitem
\bibitem[Peters \latin{et~al.}(2015)Peters, Sagar, Lantermann, and
  el~Moctar]{peters2015numerical}
Peters,~A.; Sagar,~H.; Lantermann,~U.; el~Moctar,~O. Numerical modelling and
  prediction of cavitation erosion. \emph{Wear} \textbf{2015}, \emph{338},
  189--201\relax
\mciteBstWouldAddEndPuncttrue
\mciteSetBstMidEndSepPunct{\mcitedefaultmidpunct}
{\mcitedefaultendpunct}{\mcitedefaultseppunct}\relax
\EndOfBibitem
\bibitem[Herbert \latin{et~al.}(2006)Herbert, Balibar, and
  Caupin]{herbert2006cavitation}
Herbert,~E.; Balibar,~S.; Caupin,~F. Cavitation pressure in water.
  \emph{Physical Review E} \textbf{2006}, \emph{74}, 041603\relax
\mciteBstWouldAddEndPuncttrue
\mciteSetBstMidEndSepPunct{\mcitedefaultmidpunct}
{\mcitedefaultendpunct}{\mcitedefaultseppunct}\relax
\EndOfBibitem
\bibitem[Caupin and Stroock(2013)Caupin, and Stroock]{caupin2013stability}
Caupin,~F.; Stroock,~A.~D. The stability limit and other open questions on
  water at negative pressure. \emph{Liquid Polymorphism} \textbf{2013},
  \emph{152}, 51--80\relax
\mciteBstWouldAddEndPuncttrue
\mciteSetBstMidEndSepPunct{\mcitedefaultmidpunct}
{\mcitedefaultendpunct}{\mcitedefaultseppunct}\relax
\EndOfBibitem
\bibitem[Fisher(1948)]{fisher1948fracture}
Fisher,~J.~C. The fracture of liquids. \emph{J. Appl. Phys} \textbf{1948},
  \emph{19}, 1062--1067\relax
\mciteBstWouldAddEndPuncttrue
\mciteSetBstMidEndSepPunct{\mcitedefaultmidpunct}
{\mcitedefaultendpunct}{\mcitedefaultseppunct}\relax
\EndOfBibitem
\bibitem[Caupin(2005)]{caupin2005liquid}
Caupin,~F. Liquid-vapor interface, cavitation, and the phase diagram of water.
  \emph{Phys. Rev. E} \textbf{2005}, \emph{71}, 051605\relax
\mciteBstWouldAddEndPuncttrue
\mciteSetBstMidEndSepPunct{\mcitedefaultmidpunct}
{\mcitedefaultendpunct}{\mcitedefaultseppunct}\relax
\EndOfBibitem
\bibitem[Azouzi \latin{et~al.}(2013)Azouzi, Ramboz, Lenain, and
  Caupin]{azouzi2013coherent}
Azouzi,~M. E.~M.; Ramboz,~C.; Lenain,~J.-F.; Caupin,~F. A coherent picture of
  water at extreme negative pressure. \emph{Nat. Phys.} \textbf{2013},
  \emph{9}, 38--41\relax
\mciteBstWouldAddEndPuncttrue
\mciteSetBstMidEndSepPunct{\mcitedefaultmidpunct}
{\mcitedefaultendpunct}{\mcitedefaultseppunct}\relax
\EndOfBibitem
\bibitem[Briggs(1950)]{briggs1950limiting}
Briggs,~L.~J. Limiting negative pressure of water. \emph{J. Appl. Phys}
  \textbf{1950}, \emph{21}, 721--722\relax
\mciteBstWouldAddEndPuncttrue
\mciteSetBstMidEndSepPunct{\mcitedefaultmidpunct}
{\mcitedefaultendpunct}{\mcitedefaultseppunct}\relax
\EndOfBibitem
\bibitem[Caupin(2015)]{caupin2015escaping}
Caupin,~F. Escaping the no man's land: Recent experiments on metastable liquid
  water. \emph{J. Non-Cryst. Solids} \textbf{2015}, \emph{407}, 441--448\relax
\mciteBstWouldAddEndPuncttrue
\mciteSetBstMidEndSepPunct{\mcitedefaultmidpunct}
{\mcitedefaultendpunct}{\mcitedefaultseppunct}\relax
\EndOfBibitem
\bibitem[M$ø$rch(2007)]{morch2007reflections}
M$ø$rch,~K.~A. Reflections on cavitation nuclei in water. \emph{Phys. Fluids}
  \textbf{2007}, \emph{19}\relax
\mciteBstWouldAddEndPuncttrue
\mciteSetBstMidEndSepPunct{\mcitedefaultmidpunct}
{\mcitedefaultendpunct}{\mcitedefaultseppunct}\relax
\EndOfBibitem
\bibitem[Gao \latin{et~al.}(2021)Gao, Wu, and Wang]{gao2021effects}
Gao,~Z.; Wu,~W.; Wang,~B. The effects of nanoscale nuclei on cavitation.
  \emph{J. Fluid Mech.} \textbf{2021}, \emph{911}, A20\relax
\mciteBstWouldAddEndPuncttrue
\mciteSetBstMidEndSepPunct{\mcitedefaultmidpunct}
{\mcitedefaultendpunct}{\mcitedefaultseppunct}\relax
\EndOfBibitem
\bibitem[Jones \latin{et~al.}(1999)Jones, Evans, and Galvin]{jones1999bubble}
Jones,~S.; Evans,~G.; Galvin,~K. Bubble nucleation from gas cavities—a
  review. \emph{Adv. Colloid Interface Sci.} \textbf{1999}, \emph{80},
  27--50\relax
\mciteBstWouldAddEndPuncttrue
\mciteSetBstMidEndSepPunct{\mcitedefaultmidpunct}
{\mcitedefaultendpunct}{\mcitedefaultseppunct}\relax
\EndOfBibitem
\bibitem[Epstein and Plesset(1950)Epstein, and Plesset]{epstein1950stability}
Epstein,~P.~S.; Plesset,~M.~S. On the stability of gas bubbles in liquid-gas
  solutions. \emph{J. Chem. Phys.} \textbf{1950}, \emph{18}, 1505--1509\relax
\mciteBstWouldAddEndPuncttrue
\mciteSetBstMidEndSepPunct{\mcitedefaultmidpunct}
{\mcitedefaultendpunct}{\mcitedefaultseppunct}\relax
\EndOfBibitem
\bibitem[Lohse and Zhang(2015)Lohse, and Zhang]{lohse2015surface}
Lohse,~D.; Zhang,~X. Surface nanobubbles and nanodroplets. \emph{Rev. Mod.
  Phys.} \textbf{2015}, \emph{87}, 981--1035\relax
\mciteBstWouldAddEndPuncttrue
\mciteSetBstMidEndSepPunct{\mcitedefaultmidpunct}
{\mcitedefaultendpunct}{\mcitedefaultseppunct}\relax
\EndOfBibitem
\bibitem[Tan \latin{et~al.}(2021)Tan, An, and Ohl]{tan2021stability}
Tan,~B.~H.; An,~H.; Ohl,~C.-D. Stability of surface and bulk nanobubbles.
  \emph{Curr. Opin. Colloid Interface Sci} \textbf{2021}, \emph{53},
  101428\relax
\mciteBstWouldAddEndPuncttrue
\mciteSetBstMidEndSepPunct{\mcitedefaultmidpunct}
{\mcitedefaultendpunct}{\mcitedefaultseppunct}\relax
\EndOfBibitem
\bibitem[Harvey \latin{et~al.}(1944)Harvey, Barnes, McElroy, Whiteley, Pease,
  and Cooper]{harvey1944bubble}
Harvey,~E.~N.; Barnes,~D.; McElroy,~W.~D.; Whiteley,~A.~H.; Pease,~D.;
  Cooper,~K. Bubble formation in animals. I. Physical factors. \textbf{1944},
  \relax
\mciteBstWouldAddEndPunctfalse
\mciteSetBstMidEndSepPunct{\mcitedefaultmidpunct}
{}{\mcitedefaultseppunct}\relax
\EndOfBibitem
\bibitem[Petsev \latin{et~al.}(2020)Petsev, Leal, and
  Shell]{petsev2020universal}
Petsev,~N.~D.; Leal,~L.~G.; Shell,~M.~S. Universal gas adsorption mechanism for
  flat nanobubble morphologies. \emph{Phys. Rev. Lett.} \textbf{2020},
  \emph{125}, 146101\relax
\mciteBstWouldAddEndPuncttrue
\mciteSetBstMidEndSepPunct{\mcitedefaultmidpunct}
{\mcitedefaultendpunct}{\mcitedefaultseppunct}\relax
\EndOfBibitem
\bibitem[Atchley \latin{et~al.}(1988)Atchley, Frizzell, Apfel, Holland,
  Madanshetty, and Roy]{atchley1988thresholds}
Atchley,~A.; Frizzell,~L.; Apfel,~R.; Holland,~C.; Madanshetty,~S.; Roy,~R.
  Thresholds for cavitation produced in water by pulsed ultrasound.
  \emph{Ultrasonics} \textbf{1988}, \emph{26}, 280--285\relax
\mciteBstWouldAddEndPuncttrue
\mciteSetBstMidEndSepPunct{\mcitedefaultmidpunct}
{\mcitedefaultendpunct}{\mcitedefaultseppunct}\relax
\EndOfBibitem
\bibitem[Madanshetty \latin{et~al.}(1991)Madanshetty, Roy, and
  Apfel]{madanshetty1991acoustic}
Madanshetty,~S.~I.; Roy,~R.~A.; Apfel,~R.~E. Acoustic microcavitation: Its
  active and passive acoustic detection. \emph{J. Acoust. Soc. Am}
  \textbf{1991}, \emph{90}, 1515--1526\relax
\mciteBstWouldAddEndPuncttrue
\mciteSetBstMidEndSepPunct{\mcitedefaultmidpunct}
{\mcitedefaultendpunct}{\mcitedefaultseppunct}\relax
\EndOfBibitem
\bibitem[Atchley and Prosperetti(1989)Atchley, and
  Prosperetti]{atchley1989crevice}
Atchley,~A.~A.; Prosperetti,~A. The crevice model of bubble nucleation.
  \emph{J. Acoust. Soc. Am.} \textbf{1989}, \emph{86}, 1065--1084\relax
\mciteBstWouldAddEndPuncttrue
\mciteSetBstMidEndSepPunct{\mcitedefaultmidpunct}
{\mcitedefaultendpunct}{\mcitedefaultseppunct}\relax
\EndOfBibitem
\bibitem[Borkent \latin{et~al.}(2009)Borkent, Gekle, Prosperetti, and
  Lohse]{borkent2009nucleation}
Borkent,~B.~M.; Gekle,~S.; Prosperetti,~A.; Lohse,~D. Nucleation threshold and
  deactivation mechanisms of nanoscopic cavitation nuclei. \emph{Phys. Fluids}
  \textbf{2009}, \emph{21}\relax
\mciteBstWouldAddEndPuncttrue
\mciteSetBstMidEndSepPunct{\mcitedefaultmidpunct}
{\mcitedefaultendpunct}{\mcitedefaultseppunct}\relax
\EndOfBibitem
\bibitem[De~la Fuente \latin{et~al.}(2013)De~la Fuente, Gonz{\'a}lez-Barrio,
  Navarro, Pab{\'o}n, Palacio, and Mascaraque]{de2013surface}
De~la Fuente,~O.~R.; Gonz{\'a}lez-Barrio,~M.; Navarro,~V.; Pab{\'o}n,~B.;
  Palacio,~I.; Mascaraque,~A. Surface defects and their influence on surface
  properties. \emph{J. Phys. Condens. Matter} \textbf{2013}, \emph{25},
  484008\relax
\mciteBstWouldAddEndPuncttrue
\mciteSetBstMidEndSepPunct{\mcitedefaultmidpunct}
{\mcitedefaultendpunct}{\mcitedefaultseppunct}\relax
\EndOfBibitem
\bibitem[Siretanu \latin{et~al.}(2016)Siretanu, van~den Ende, and
  Mugele]{siretanu2016atomic}
Siretanu,~I.; van~den Ende,~D.; Mugele,~F. Atomic structure and surface defects
  at mineral-water interfaces probed by in situ atomic force microscopy.
  \emph{Nanoscale} \textbf{2016}, \emph{8}, 8220--8227\relax
\mciteBstWouldAddEndPuncttrue
\mciteSetBstMidEndSepPunct{\mcitedefaultmidpunct}
{\mcitedefaultendpunct}{\mcitedefaultseppunct}\relax
\EndOfBibitem
\bibitem[Lum \latin{et~al.}(1999)Lum, Chandler, and
  Weeks]{lum1999hydrophobicity}
Lum,~K.; Chandler,~D.; Weeks,~J.~D. Hydrophobicity at small and large length
  scales. 1999\relax
\mciteBstWouldAddEndPuncttrue
\mciteSetBstMidEndSepPunct{\mcitedefaultmidpunct}
{\mcitedefaultendpunct}{\mcitedefaultseppunct}\relax
\EndOfBibitem
\bibitem[Huang \latin{et~al.}(2003)Huang, Margulis, and
  Berne]{huang2003dewetting}
Huang,~X.; Margulis,~C.~J.; Berne,~B.~J. Dewetting-induced collapse of
  hydrophobic particles. \emph{Proc. Natl. Acad. Sci. U. S. A.} \textbf{2003},
  \emph{100}, 11953--11958\relax
\mciteBstWouldAddEndPuncttrue
\mciteSetBstMidEndSepPunct{\mcitedefaultmidpunct}
{\mcitedefaultendpunct}{\mcitedefaultseppunct}\relax
\EndOfBibitem
\bibitem[Sharma and Debenedetti(2012)Sharma, and
  Debenedetti]{sharma2012evaporation}
Sharma,~S.; Debenedetti,~P.~G. Evaporation rate of water in hydrophobic
  confinement. \emph{Proc. Natl. Acad. Sci. U. S. A.} \textbf{2012},
  \emph{109}, 4365--4370\relax
\mciteBstWouldAddEndPuncttrue
\mciteSetBstMidEndSepPunct{\mcitedefaultmidpunct}
{\mcitedefaultendpunct}{\mcitedefaultseppunct}\relax
\EndOfBibitem
\bibitem[Abascal and Vega(2005)Abascal, and Vega]{tip4p_water}
Abascal,~J. L.~F.; Vega,~C. {A general purpose model for the condensed phases
  of water: TIP4P/2005}. \emph{J. Chem. Phys.} \textbf{2005}, \emph{123},
  234505\relax
\mciteBstWouldAddEndPuncttrue
\mciteSetBstMidEndSepPunct{\mcitedefaultmidpunct}
{\mcitedefaultendpunct}{\mcitedefaultseppunct}\relax
\EndOfBibitem
\bibitem[Vanommeslaeghe \latin{et~al.}(2010)Vanommeslaeghe, Hatcher, Acharya,
  Kundu, Zhong, Shim, Darian, Guvench, Lopes, Vorobyov, and
  Mackerell]{vanommeslaeghe2010charmm}
Vanommeslaeghe,~K.; Hatcher,~E.; Acharya,~C.; Kundu,~S.; Zhong,~S.; Shim,~J.;
  Darian,~E.; Guvench,~O.; Lopes,~P.; Vorobyov,~I.; Mackerell,~A. CHARMM
  general force field: A force field for drug-like molecules compatible with
  the CHARMM all-atom additive biological force fields. \emph{J. Comput. Chem.}
  \textbf{2010}, \emph{31}, 671--690\relax
\mciteBstWouldAddEndPuncttrue
\mciteSetBstMidEndSepPunct{\mcitedefaultmidpunct}
{\mcitedefaultendpunct}{\mcitedefaultseppunct}\relax
\EndOfBibitem
\bibitem[Šako \latin{et~al.}(2023)Šako, Staniscia, Schneck, Netz, and
  Kanduč]{10.1093/pnasnexus/pgad190}
Šako,~M.; Staniscia,~F.; Schneck,~E.; Netz,~R.~R.; Kanduč,~M. {Conditions for
  the stable adsorption of lipid monolayers to solid surfaces}. \emph{PNAS
  Nexus} \textbf{2023}, \emph{2}, pgad190\relax
\mciteBstWouldAddEndPuncttrue
\mciteSetBstMidEndSepPunct{\mcitedefaultmidpunct}
{\mcitedefaultendpunct}{\mcitedefaultseppunct}\relax
\EndOfBibitem
\bibitem[Strong and Whitesides(1988)Strong, and
  Whitesides]{strong1988structures}
Strong,~L.; Whitesides,~G.~M. Structures of self-assembled monolayer films of
  organosulfur compounds adsorbed on gold single crystals: electron diffraction
  studies. \emph{Langmuir} \textbf{1988}, \emph{4}, 546--558\relax
\mciteBstWouldAddEndPuncttrue
\mciteSetBstMidEndSepPunct{\mcitedefaultmidpunct}
{\mcitedefaultendpunct}{\mcitedefaultseppunct}\relax
\EndOfBibitem
\bibitem[Chidsey and Loiacono(1990)Chidsey, and Loiacono]{chidsey1990chemical}
Chidsey,~C.~E.; Loiacono,~D.~N. Chemical functionality in self-assembled
  monolayers: structural and electrochemical properties. \emph{Langmuir}
  \textbf{1990}, \emph{6}, 682--691\relax
\mciteBstWouldAddEndPuncttrue
\mciteSetBstMidEndSepPunct{\mcitedefaultmidpunct}
{\mcitedefaultendpunct}{\mcitedefaultseppunct}\relax
\EndOfBibitem
\bibitem[Fenter \latin{et~al.}(1997)Fenter, Eberhardt, Liang, and
  Eisenberger]{fenter1997epitaxy}
Fenter,~P.; Eberhardt,~A.; Liang,~K.; Eisenberger,~P. Epitaxy and chainlength
  dependent strain in self-assembled monolayers. \emph{J. Chem. Phys.}
  \textbf{1997}, \emph{106}, 1600--1608\relax
\mciteBstWouldAddEndPuncttrue
\mciteSetBstMidEndSepPunct{\mcitedefaultmidpunct}
{\mcitedefaultendpunct}{\mcitedefaultseppunct}\relax
\EndOfBibitem
\bibitem[Abraham \latin{et~al.}(2015)Abraham, Murtola, Schulz, Páll, Smith,
  Hess, and Lindahl]{ABRAHAM201519}
Abraham,~M.~J.; Murtola,~T.; Schulz,~R.; Páll,~S.; Smith,~J.~C.; Hess,~B.;
  Lindahl,~E. GROMACS: High performance molecular simulations through
  multi-level parallelism from laptops to supercomputers. \emph{SoftwareX}
  \textbf{2015}, \emph{1-2}, 19--25\relax
\mciteBstWouldAddEndPuncttrue
\mciteSetBstMidEndSepPunct{\mcitedefaultmidpunct}
{\mcitedefaultendpunct}{\mcitedefaultseppunct}\relax
\EndOfBibitem
\bibitem[Bussi \latin{et~al.}(2007)Bussi, Donadio, and Parrinello]{v-rescale}
Bussi,~G.; Donadio,~D.; Parrinello,~M. Canonical sampling through velocity
  rescaling. \emph{J. Chem. Phys.} \textbf{2007}, \emph{126}, 014101\relax
\mciteBstWouldAddEndPuncttrue
\mciteSetBstMidEndSepPunct{\mcitedefaultmidpunct}
{\mcitedefaultendpunct}{\mcitedefaultseppunct}\relax
\EndOfBibitem
\bibitem[Bernetti and Bussi(2020)Bernetti, and Bussi]{c-rescale}
Bernetti,~M.; Bussi,~G. Pressure control using stochastic cell rescaling.
  \emph{J. Chem. Phys.} \textbf{2020}, \emph{153}, 114107\relax
\mciteBstWouldAddEndPuncttrue
\mciteSetBstMidEndSepPunct{\mcitedefaultmidpunct}
{\mcitedefaultendpunct}{\mcitedefaultseppunct}\relax
\EndOfBibitem
\bibitem[H{\"a}nggi \latin{et~al.}(1990)H{\"a}nggi, Talkner, and
  Borkovec]{hanggi1990reaction}
H{\"a}nggi,~P.; Talkner,~P.; Borkovec,~M. Reaction-rate theory: fifty years
  after Kramers. \emph{Rev. Mod. Phys.} \textbf{1990}, \emph{62}, 251\relax
\mciteBstWouldAddEndPuncttrue
\mciteSetBstMidEndSepPunct{\mcitedefaultmidpunct}
{\mcitedefaultendpunct}{\mcitedefaultseppunct}\relax
\EndOfBibitem
\bibitem[Kandu{\v c} \latin{et~al.}(2020)Kandu{\v c}, Schneck, Loche, Jansen,
  Schenk, and Netz]{Kanduc10733}
Kandu{\v c},~M.; Schneck,~E.; Loche,~P.; Jansen,~S.; Schenk,~H.~J.; Netz,~R.~R.
  Cavitation in lipid bilayers poses strict negative pressure stability limit
  in biological liquids. \emph{Proc. Natl. Acad. Sci. U. S. A.} \textbf{2020},
  \emph{117}, 10733--10739\relax
\mciteBstWouldAddEndPuncttrue
\mciteSetBstMidEndSepPunct{\mcitedefaultmidpunct}
{\mcitedefaultendpunct}{\mcitedefaultseppunct}\relax
\EndOfBibitem
\bibitem[Boucher \latin{et~al.}(2007)Boucher, Jo{\'o}s, Zuckermann, and
  Fournier]{boucher2007pore}
Boucher,~P.-A.; Jo{\'o}s,~B.; Zuckermann,~M.~J.; Fournier,~L. Pore formation in
  a lipid bilayer under a tension ramp: modeling the distribution of rupture
  tensions. \emph{Biophys. J.} \textbf{2007}, \emph{92}, 4344--4355\relax
\mciteBstWouldAddEndPuncttrue
\mciteSetBstMidEndSepPunct{\mcitedefaultmidpunct}
{\mcitedefaultendpunct}{\mcitedefaultseppunct}\relax
\EndOfBibitem
\bibitem[Šako \latin{et~al.}(2025)Šako, Netz, and Kanduč]{sako2024impact}
Šako,~M.; Netz,~R.~R.; Kanduč,~M. Impact of Nanoscopic Impurity Aggregates on
  Cavitation in Water. \emph{Surf. Interfaces.} \textbf{2025}, \emph{xx},
  xx\relax
\mciteBstWouldAddEndPuncttrue
\mciteSetBstMidEndSepPunct{\mcitedefaultmidpunct}
{\mcitedefaultendpunct}{\mcitedefaultseppunct}\relax
\EndOfBibitem
\bibitem[Fischler and Bolles(1981)Fischler, and Bolles]{ransac}
Fischler,~M.~A.; Bolles,~R.~C. Random Sample Consensus: A Paradigm for Model
  Fitting with Applications to Image Analysis and Automated Cartography.
  \emph{Commun. ACM} \textbf{1981}, \emph{24}, 381–395\relax
\mciteBstWouldAddEndPuncttrue
\mciteSetBstMidEndSepPunct{\mcitedefaultmidpunct}
{\mcitedefaultendpunct}{\mcitedefaultseppunct}\relax
\EndOfBibitem
\bibitem[Menzl \latin{et~al.}(2016)Menzl, Gonzalez, Geiger, Caupin, Abascal,
  Valeriani, and Dellago]{menzl2016molecular}
Menzl,~G.; Gonzalez,~M.~A.; Geiger,~P.; Caupin,~F.; Abascal,~J.~L.;
  Valeriani,~C.; Dellago,~C. Molecular mechanism for cavitation in water under
  tension. \emph{Proc. Natl. Acad. Sci. U. S. A.} \textbf{2016}, \emph{113},
  13582--13587\relax
\mciteBstWouldAddEndPuncttrue
\mciteSetBstMidEndSepPunct{\mcitedefaultmidpunct}
{\mcitedefaultendpunct}{\mcitedefaultseppunct}\relax
\EndOfBibitem
\bibitem[Volmer(1929)]{volmer1929keimbildung}
Volmer,~M. {\"U}ber keimbildung und keimwirkung als spezialf{\"a}lle der
  heterogenen katalyse. \emph{Z. Elektrochem. Angew. Phys. Chem.}
  \textbf{1929}, \emph{35}, 555--561\relax
\mciteBstWouldAddEndPuncttrue
\mciteSetBstMidEndSepPunct{\mcitedefaultmidpunct}
{\mcitedefaultendpunct}{\mcitedefaultseppunct}\relax
\EndOfBibitem
\bibitem[Carlson \latin{et~al.}(2021)Carlson, Becker, Brünig, Ataka, Cruz, Yu,
  Tang, Kanduč, Haag, Heberle, Makki, and Netz]{carlson2021hydrophobicity}
Carlson,~S.; Becker,~M.; Brünig,~F.~N.; Ataka,~K.; Cruz,~R.; Yu,~L.; Tang,~P.;
  Kanduč,~M.; Haag,~R.; Heberle,~J.; Makki,~H.; Netz,~R.~R. Hydrophobicity of
  self-assembled monolayers of alkanes: fluorination, density, roughness, and
  Lennard-Jones cutoffs. \emph{Langmuir} \textbf{2021}, \emph{37},
  13846--13858\relax
\mciteBstWouldAddEndPuncttrue
\mciteSetBstMidEndSepPunct{\mcitedefaultmidpunct}
{\mcitedefaultendpunct}{\mcitedefaultseppunct}\relax
\EndOfBibitem
\bibitem[Loche \latin{et~al.}(2024)Loche, Kandu{\v{c}}, Schneck, and
  Netz]{loche2024water}
Loche,~P.; Kandu{\v{c}},~M.; Schneck,~E.; Netz,~R.~R. Water cavitation results
  from the kinetic competition of bulk, surface and surface-defect nucleation
  events. \emph{arXiv preprint arXiv:2410.17626} \textbf{2024}, \relax
\mciteBstWouldAddEndPunctfalse
\mciteSetBstMidEndSepPunct{\mcitedefaultmidpunct}
{}{\mcitedefaultseppunct}\relax
\EndOfBibitem
\bibitem[Roger and Cabane(2012)Roger, and Cabane]{roger2012hydrophobic}
Roger,~K.; Cabane,~B. Why are hydrophobic/water interfaces negatively charged?
  \emph{Angew. Chem., Int. Ed. Engl.} \textbf{2012}, \emph{51}, 5625\relax
\mciteBstWouldAddEndPuncttrue
\mciteSetBstMidEndSepPunct{\mcitedefaultmidpunct}
{\mcitedefaultendpunct}{\mcitedefaultseppunct}\relax
\EndOfBibitem
\bibitem[Maali \latin{et~al.}(2017)Maali, Boisgard, Chraibi, Zhang, Kellay, and
  W{\"u}rger]{maali2017viscoelastic}
Maali,~A.; Boisgard,~R.; Chraibi,~H.; Zhang,~Z.; Kellay,~H.; W{\"u}rger,~A.
  Viscoelastic drag forces and crossover from no-slip to slip boundary
  conditions for flow near air-water interfaces. \emph{Phys. Rev. Lett.}
  \textbf{2017}, \emph{118}, 084501\relax
\mciteBstWouldAddEndPuncttrue
\mciteSetBstMidEndSepPunct{\mcitedefaultmidpunct}
{\mcitedefaultendpunct}{\mcitedefaultseppunct}\relax
\EndOfBibitem
\bibitem[Uematsu \latin{et~al.}(2019)Uematsu, Bonthuis, and
  Netz]{uematsu2019impurity}
Uematsu,~Y.; Bonthuis,~D.~J.; Netz,~R.~R. Impurity effects at hydrophobic
  surfaces. \emph{Curr. Opin. Electrochem.} \textbf{2019}, \emph{13},
  166--173\relax
\mciteBstWouldAddEndPuncttrue
\mciteSetBstMidEndSepPunct{\mcitedefaultmidpunct}
{\mcitedefaultendpunct}{\mcitedefaultseppunct}\relax
\EndOfBibitem
\bibitem[Zhang \latin{et~al.}(2021)Zhang, Yang, Ngo, Guo, Wen, Wang, Zhang, and
  Long]{zhang2021critical}
Zhang,~X.; Yang,~Y.; Ngo,~H.~H.; Guo,~W.; Wen,~H.; Wang,~X.; Zhang,~J.;
  Long,~T. A critical review on challenges and trend of ultrapure water
  production process. \emph{Sci. Total Environ.} \textbf{2021}, \emph{785},
  147254\relax
\mciteBstWouldAddEndPuncttrue
\mciteSetBstMidEndSepPunct{\mcitedefaultmidpunct}
{\mcitedefaultendpunct}{\mcitedefaultseppunct}\relax
\EndOfBibitem
\bibitem[Schenk \latin{et~al.}(2015)Schenk, Steppe, and
  Jansen]{schenk2015nanobubbles}
Schenk,~H.~J.; Steppe,~K.; Jansen,~S. Nanobubbles: a new paradigm for
  air-seeding in xylem. \emph{Trends Plant Sci.} \textbf{2015}, \emph{20},
  199--205\relax
\mciteBstWouldAddEndPuncttrue
\mciteSetBstMidEndSepPunct{\mcitedefaultmidpunct}
{\mcitedefaultendpunct}{\mcitedefaultseppunct}\relax
\EndOfBibitem
\end{mcitethebibliography}

\end{document}